\documentclass[useAMS,usenatbib]{mn2e}
\usepackage[normalem]{ulem}
\usepackage[dvips]{graphicx, color}

\title[Energetics of the accretion flow in the VHS]
  {Tracking the energetics of the non-thermal disc-corona-jet in the
very high state GX~$339-4$ }
\author[A. Kubota \& C. Done]
  {A.~Kubota,$^1$\thanks{Affiliated to Durham University.}
  C.~Done,$^2$\\
  $^1$Department of Electronic Information Systems, Shibaura Institute of Technology, 307 Fukasaku, Minuma-ku, \\
  Saitama-shi, Saitam 337-8570, Japan; aya@shibaura-it.ac.jp\\
  $^2$Department of Physics, University of Durham, South Road, Durham, DH1 3LE, England; chris.done@durham.ac.uk}
\date{Released 2014 Xxxxx XX}

\pagerange{\pageref{firstpage}--\pageref{lastpage}} \pubyear{2002}

\def\LaTeX{L\kern-.36em\raise.3ex\hbox{a}\kern-.15em
    T\kern-.1667em\lower.7ex\hbox{E}\kern-.125emX}

\begin{document}

\label{firstpage}

\maketitle

\begin{abstract}

The dramatic hard-soft spectral transition in Black Hole Binaries is important as it is associated with the collapse of the jet and with the strongest low frequency QPOs. These transition spectra (intermediate and very high state: VHS) are complex, with soft but distinctly non-thermal Comptonisation which merges smoothly into the disc emission. Here we develop a physical model for the accretion flow which can accommodate all these features, with an outer standard disc, which can make a transition to an energetically coupled disc-corona region, and make a further transition to a hot inner flow which can be radiatively inefficient if required. The code explicitly uses fully relativistic emissivity (Novikov-Thorne), and all Comptonisation is calculated with a hybrid (thermal and non-thermal) electron distribution.
We fit this to a VHS spectrum from GX~$339-4$. We show that the complex continuum curvature produced by a hybrid electron distribution is enough to remove the strong constraint on black hole spin derived from reflection using simpler Comptonisation models. More fundamentally, we show that the VHS cannot be fit with the same Novikov-Thorne emissivity which can fit the disc dominated spectrum but instead requires that the inner flow is somewhat radiatively inefficient. This is consistent with an accretion powered jet, but simultaneous radio data show that the jet has already collapsed at the time of our data. Instead, it could point to truncation of the inner flow at radii larger than the innermost stable circular orbit, as predicted by the Lense-Thirring QPO models.

\end{abstract}


\section{Introduction }

The two best known black hole binary states are the low/hard and
high/soft states. The low/hard state is dominated by hard (photon
index $\Gamma<2$) Comptonised emission from a hot, optically thin flow
whereas the high/soft state is instead dominated by soft, blackbody
emission from a cool, optically thick disc. The transition between
these two states is generally interpreted in terms of the truncated
disc model, where in the low/hard state the inner thin, cool,
radiatively efficient disc evaporates into a geometrically thick, hot,
advection dominated flow (e.g., Esin, McClintock, \& Narayan
1997). Spectra around the transition are rather rare since the
transitions are fast, but these (hard and soft) intermediate states
(HIMS/SIMS) show both a strong disc and strong but soft ($\Gamma\sim
2.5-3$) Comptonised tail. There is also a distinct branch, the
  very high state (VHS), on the
hardness-intensity diagram at the highest luminosity where the spectral shape is similarly a composite of a strong
disc and strong, steep Comptonised spectrum~\citep{rm06, dgk07}.

These transition spectra are of especial importance as they are
associated with dramatic changes in the radio emission from the jet.
The steady compact jet seen in the low/hard state 
(Gallo, Fender \& Pooley 2003; or Corbel et al 2013a for an up to date version)
collapses as the
source makes a transition to the disc dominated states. 
This collapse
takes place via discrete plasma ejection events which can impact into
the previous jet material, producing optically thin synchrotron
emission from internal shocks \citep{fender04}.
Extrapolating back from the observed proper motion of the shocked
radio plasma indicates that the initial ejection takes place around
the time where the fast X-ray variability properties change
dramatically (e.g. Miller-Jones et al 2012). The ubiquitous broad
band noise seen in the X-ray variability in the low/hard state
suddenly collapses, correlated with a more subtle change in the low
frequency Quasi-Periodic Oscillation (QPO) from type C to B, and an
even more subtle spectral softening (from HIMS to SIMS). It is very
tempting to associate this very obvious change in timing signature with the
ejection event (jet line, Fender et al.~2004), but this is probably not a
causal connection as some 
objects with good monitoring campaigns show that the 
radio collapses before the change in timing (GX~$339-4$: Fender 2009; 
MAXI~J$1659-152$: van der Horst et al 2013).

Nonetheless, the steady compact jet is normally seen together with the 
type C QPO. The current best model for this  QPO involves Lense-Thirring
(vertical) precession of the entire hot inner flow region within the
truncated disc (Ingram, Done \& Fragile 2009), so this associates
the steady jet with a geometrically thick, hot inner flow. However, 
testing whether this truncated disc/hot inner flow geometry really
does apply in the HIMS/VHS spectra is not easy. 

The inner edge of the thin disc can be tracked in the disc dominated states, and
is consistent with extending down to a fixed inner radius despite
large changes in mass accretion rate \citep{kme01,km04,dgk07}. 
However, the HIMS/VHS are characterised by strong
soft Comptonisation of the disc emission, which can distort the
observed disc emission in two ways. Firstly, strong Comptonisation implies that not all of
the accretion energy is dissipated in the disc.  In a disc-corona
geometry, this means that the underlying disc is cooler and less
luminous than expected \citep{poutanen96}.  Secondly, the
Comptonisation process itself removes photons from the disc spectrum
to scatter them into the Compton spectrum. Thus there is coupling of
both energy and photon number between the disc and Comptonisation
region.  In the studies of \citet{kd04}, \citet{dkbbfth} and
\citet{tamura12} (hereafter T12), we assumed that this coupling was
between a disc and thermal Compton component.  However, the strong
Compton tail in the VHS/HIMS is often seen to have a composite shape,
requiring a non-thermal electron distribution as well as thermal
electrons \citep{zdziarski01, gd03, Hjalmarsdotter15}. In our previous work, we assumed
that this non-thermal emission had negligible impact on the energy and
photon coupling between the disc and corona. This is not necessarily
true.

In this paper, we build a new model `{\sc diskEQ}' to explore the
geometry and energetics of these complex HIMS/VHS.  We use
the Novikov-Thorne thin disc emissivity to specify the local energy
release. At large radii we assume that this emission is dissipated as
a (colour temperature corrected) blackbody as in a standard disc. We
allow this to make a transition in the inner region to an 
energetically coupled disc-corona, where the coronal electrons can
have a hybrid (thermal and non-thermal) electron distribution,
modeled using the {\sc eqpair} code \citep{coppi92}. We then allow
the disc-corona to make a further transition to a hot inner flow,
where again the electrons can have a hybrid electron distribution.

We use this to fit the VHS/HIMS spectrum of a black hole binary
GX~$339-4$ observed with Suzaku, and compare this with a disc
dominated spectrum seen from the same object three days later with
XMM-Newton and RXTE in order to get the best possible constraints on
any change in geometry and energetics between these two states.

In Section~2, we describe observations of the VHS with Suzaku and the
high/soft state with XMM and RXTE.  Details of the new model are given
in Section~3. The high/soft state spectrum is fit in Section~4. This
is disc dominated, so it constrains black hole spin. This is rather
low for our assumed system parameters, in conflict with the very high
black hole spin derived from a recent re-evaluation of the iron line
profile in multiple datasets, including the VHS Suzaku data studied
here \citep{ludlam15}. We fit the VHS spectra firstly
with simple models in Section~5 and show that the high spin/high
reflected fraction derived by Ludlam et al (2015) is degenerate with
the continuum complexity which comes from allowing a hybrid (rather
than purely power law) electron distribution. We then fit the VHS
spectra with our new disc model, to show that it requires an inner
flow which is underluminous compared to the mass accretion rate
inferred through the outer disc. While this might signal that a
substantial fraction of the accretion luminosity is used to power the
jet, the absence of strong radio emission during this observation
makes this unlikely. We conclude that either the flow has not yet
reached steady state from the rapid rise to outburst and/or that the
Novikov-Thorne accretion emissivity derived for a thin disc is not
appropriate for the more complex accretion flow geometry considered
here.

\section{observation and data reduction}

As described in T12, the 2006/2007 outburst of GX~$339-4$ was its
brightest outburst since the launch of RXTE in 1995.  The light curves
and hardness ratio of this outburst (Figure 1 of T12) showed that the
state transition occurred around MJD$\simeq$54140--54150, and that the
Suzaku observation was performed just before the peak of the RXTE
count rate (MJD 54143.2--54146.2 ; 2007 February 12th 05:33:31
to 15th 04:48:26).  Both the QPO (type C, increasing from 4.3-5.5~Hz)
and spectral properties ($\Gamma$ increasing from 2.6-2.7)
characterize the source as being in the VHS during the Suzaku
observation (T12).

Data reduction is described in T12, but we redo this  by {\sc
aepipeline (version 1.1.0) }implemented in {\sc heasoft~6.16 } using the
same selection but with the latest calibration files released on 2014
June 24th. As before, we mitigate pileup in the XIS by using only XIS0
data, and extract events from a circular annulus from 7$^\prime$--3$^\prime$, to
exclude the core \citep{yamada09}. Further excluding events with
telemetry saturation gave an exposure time of 2.83~ks out of the total
12.4~ks observation. We use the XIS0 data from 0.7--1.5~keV and 2.3--10~keV,
including 1\% systematic error on each energy bin.

We similarly reprocessed the Hard X-ray Detector (HXD;
\citealt{kokubun07}) data from both PIN and GSO data with {\sc
aepipeline} with their standard criteria. Following T12, both the non
X-ray background (NXB) and the cosmic X-ray background (CXB) were
subtracted from the PIN spectrum, and only NXB was subtracted from the
GSO spectrum.  For the spectral analyses, we used 12--70~keV PIN data
with the response matrix of {\sc ae\_hxd\_pinxinome3\_20080129.rsp},
and used 50--200~keV GSO data with the response matrixes of {\sc
ae\_hxd\_gsoxinom\_20100524.rsp} and {\sc
ae\_hxd\_gsoxinom\_crab\_20100526.arf} \citep{yamada11}.  Systematic
errors of 1\% were added to each energy bin for the PIN spectrum.  We
also added 0.5\% to the GSO background spectrum to consider the effect of a systematic
uncertainty of up to 3\% in the GSO background.  The cross
normalization of PIN and GSO to XIS0 are fixed at
1.16\footnote{ftp://legacy.gsfc.nasa.gov/suzaku/doc/xrt/suzakumemo-2008-06.pdf},
since the observation was performed at the XIS nominal position.

We use the high/soft state data from XMM and RXTE taken at the peak of
the RXTE/ASM count rate (MJD 54150.01--54150.20; 2007 February
19th 00:20:54 to 04:43:28). The XMM-Newton data are in burst mode,
with net exposure of 448~s, while the contemporaneous RXTE data have
net exposure of 3.23~ks (see Kolehmainen, Done, \& D{\'{\i}}az Trigo
2011). We use the energy range 0.6--2.0~keV and 2.4--10~keV for the XMM/PN,
7--30~keV for the RXTE/PCA, and 30--150 keV for the RXTE/HEXTE.  We
use systematic errors of 1\% to the each energy bin for the XMM/PN
spectrum. This is lower than the 2\% used by Kolehmainen et al. (2011)
as we do not include the RXTE data in the energy range of 3-7~keV
where the cross-calibration errors are large (Kolehmainen et
al. 2011).

\section{{\sc diskEQ} Model}

\begin{figure*}
\includegraphics[width=58mm]{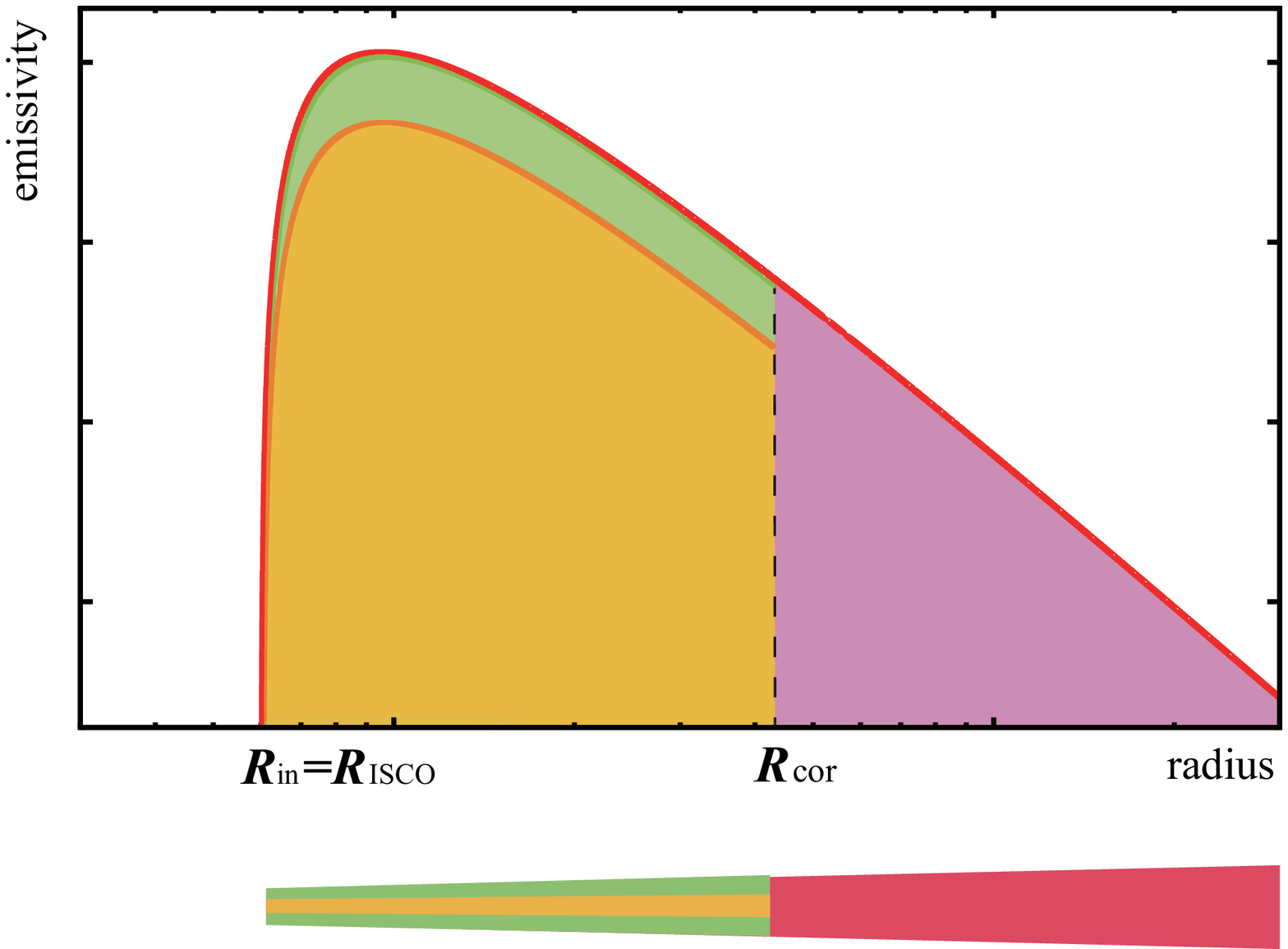}
\includegraphics[width=58mm]{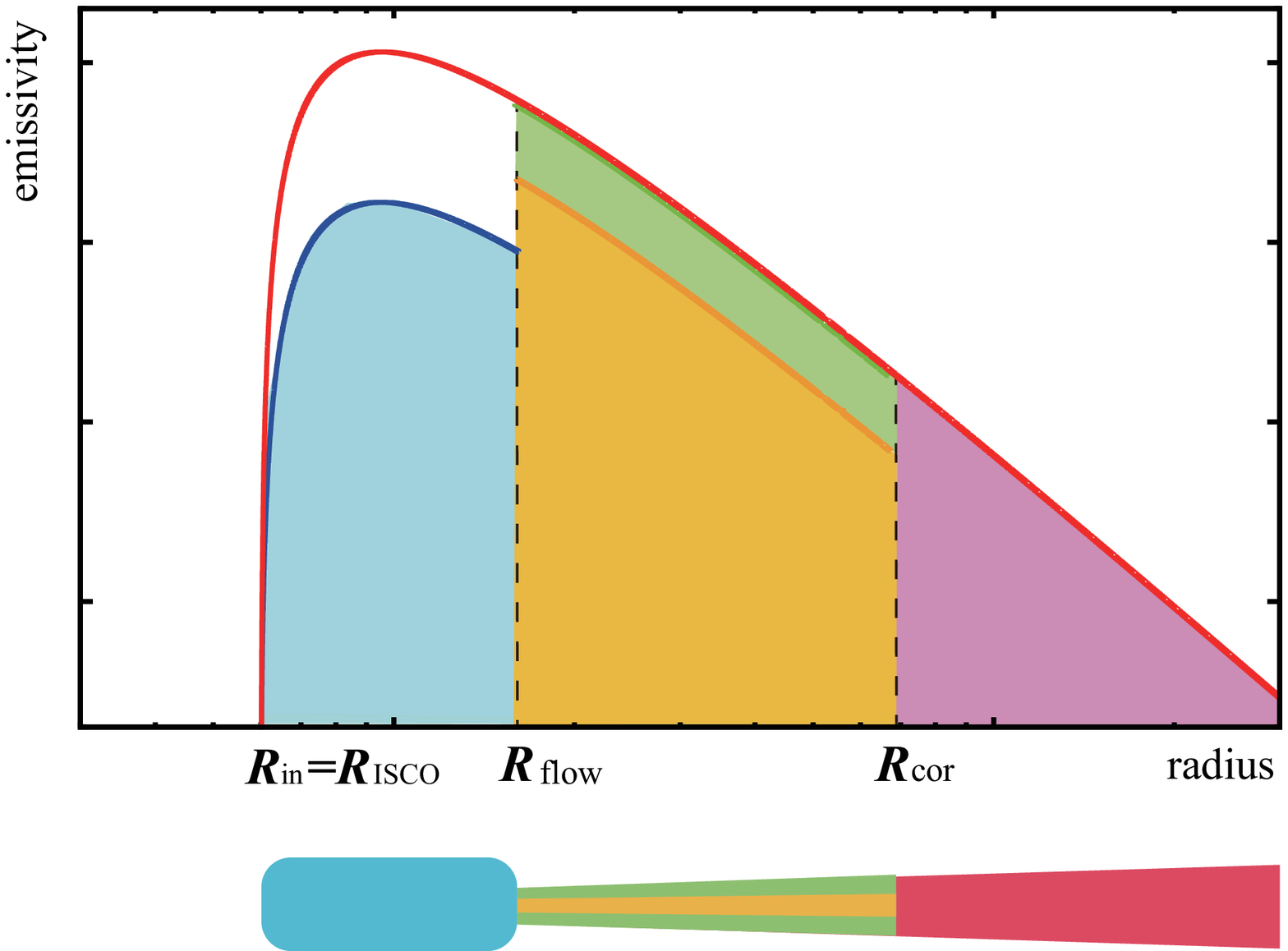}
\includegraphics[width=58mm]{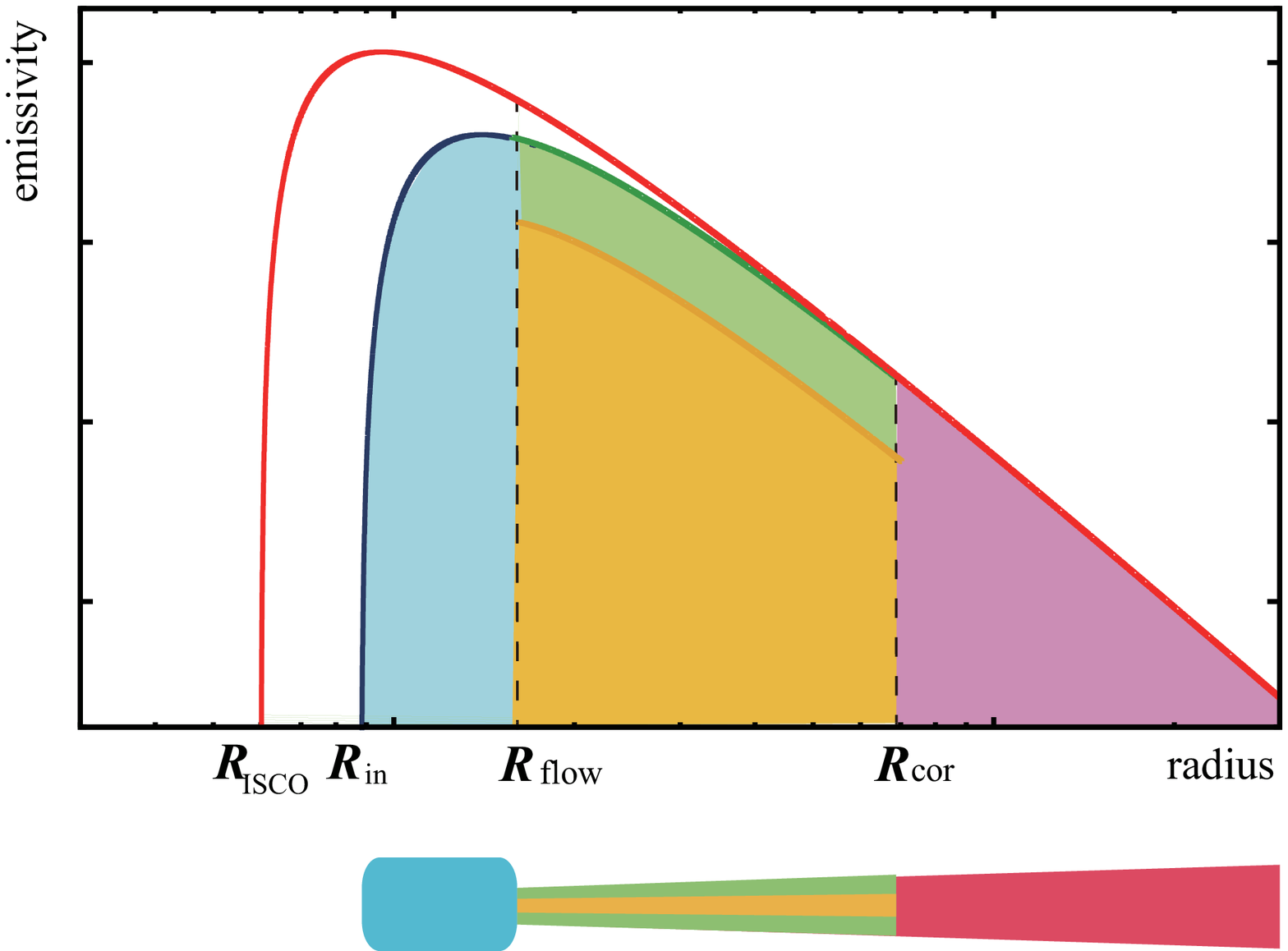}
 \caption{schematic picture of accretion disc structure and local emissivity of disc without inner flow (left), 
  disc with inefficient inner flow (middle), 
 and  disc with stress-free truncated inner flow (right). 
 Novikov-Thorne model,  black body emission from the outer disc, black body emission from the middle part, {\sc eqpair} from the middle part, and the comptonised emission from the inner plasma are shown with 
 solid red lines, magenta, orange, green, and blue, respectively. }
\label{fig:nt}
\end{figure*}

The energetically coupled disc and corona model, {\sc dkbbfth}
\citep{dkbbfth}, used to fit these data in T12 is based on the {\sc
diskbb} \citep{diskbb} emissivity i.e. assumes that the local energy
release is proportional to $R^{-3}$ in the disc over all radii from
$R_{\rm out}$ down to $R_{\rm in}$. This gives a temperature
distribution $T(r)=T(R_{\rm in}) (R/R_{\rm in})^{-3/4}$ if it thermalises to a
blackbody, so it is parameterised by an inner disc temperature,
$T(R_{\rm in})$, and radius $R_{\rm in}$.  We upgrade this to use the fully
general relativistic Novikov-Thorne emissivity \citep{nt73}, which
includes the full stress-free inner boundary condition.  The free
parameters are then the three physical parameters of mass, mass
accretion rate and black hole spin.

As in {\sc dkbbfth}, the model assumes that this energy is dissipated
as an optically thick blackbody from $R_{\rm out}$ to $R_{\rm cor}$. However, the optically thick disc emission is now explicitly
corrected for electron scattering in the disc via a self consistently
calculated colour temperature correction 
\citep{optxagn}, rather than having this as an additional free parameter.  We calculate the 
colour tempearture corrected blackbody flux on a radial grid with at least 20 points per decade.

We then use the {\sc eqpair}\footnote{http://www.astro.yale.edu/coppi/eqpair/eqpap4.ps} code
(e.g.,Coppi~1992) to describe the Comptonisation, so that the electron
distribution can be a mix of thermal and non-thermal (hybrid).  The
spectral shape of {\sc eqpair} is parameterised using the ratio of
power injected into the electrons (hard power, $L_{\rm h}$) to the
luminosity of seed photons intercepted by the source (seed photon
luminosity, $L_{\rm s}$). Both these are parameterised as a compactness
i.e. a dimensionless luminosity to size ratio i.e.  $\ell=L/R \times
\sigma_{\rm T} / (m_e c^3)$, and the ratio $L_{\rm h}/L_{\rm s}=\ell_{\rm h}/\ell_{\rm s}$. 
A further
parameter sets the fraction of non-thermal acceleration compared to
the total (non-thermal plus thermal) electron power $\ell_{\rm
nth}/\ell_{\rm h}$. Any non-thermal electrons are assumed to be
injected with a power law spectrum, $Q(\gamma)\propto
\gamma^{-\Gamma_{\rm inj}}$, between $\gamma_{\rm min}$ and $\gamma_{\rm
max}$ while the thermal ones have a Maxwellian distribution.  Both
thermal and non-thermal electrons cool via Compton and Coulomb
collisions to form a self-consistent hybrid electron distribution,
$N(\gamma)$. Even completely non-thermal acceleration gives a steady
state distribution which is thermal at low energies due to the effect
of Coulomb collisions. We fix the minimum and maximum electron
Lorentz factors $\gamma_{\rm min}$ and $\gamma_{\rm max}$ at 1.3 and
1000, respectively, leaving only $\Gamma_{\rm inj}$ as the free parameter
of the non-thermal electron injection. The code includes the effects
of photon-photon pair production, so the self-consistent electron
optical depth $\int N(\gamma)\sigma_{\rm T} R d\gamma=\tau=
\tau_p+\tau_{e^{\pm}}$.

We consider several geometries for the electrons in the VHS, as shown
schematically in Figure 1.  Firstly (Fig 1a) we assume the
Comptonising corona overlies the inner portion of an untruncated inner
disc as in the {\sc dkbbfth} model but with the more flexible {\sc
  eqpair} model to describe the Comptonised emission, allowing the
electrons to be non-thermal as well as thermal.  We assume that all
the electron parameters remain constant with radius in the coupled
disc-corona region, and that the accretion power dissipated in radii
between $R_{\rm cor}$ to $R_{\rm ISCO}$ is split between the disc and
corona, with a fraction $f$ powering the corona, and the remaining
$(1-f)$ powering the underlying disc
\citep{haardt93,svensson94}. Hence the underlying disc emission is
both less luminous and cooler than a pure disc without a corona.
However, there is also reprocessing of the hard X-rays in the disc,
which adds to the seed photons, so the ratio of coronal power to seed
photon luminosity, $\ell_{\rm h}/\ell_{\rm s}$, which is the major determinant of
the shape of the spectrum is not simply $f/(1-f)$. Instead, we use
$l_{\rm h}/l_{\rm s}$ derived from the shape of the spectrum to define a new
parameter $g=\ell_{\rm h}/(\ell_{\rm s}+\ell_{\rm h})$ which includes the contribution
of reprocessed/reflected flux. The relation of $g$ to $f$ is described
in Appendix B, but they are equal only if the albedo $a=1$ in the
coupled disc-corona region, so that all the flux is reflected rather
than reprocessed. 

We calculate Comptonisation locally, splitting the coupled disc-corona
region up into multiple regions, again using a radial grid with at
least 20 points per decade.  The soft compactness in each annulus of
width $dR$ is $\ell_{\rm s}=[(1-g) L_{\rm NT}(R) 4\pi R dR/R]\times
(\sigma_{\rm T}/ m_e c^3)$, while the intrinsic hard compactness is
$\ell_{\rm h}=[ g L_{\rm NT}(R) 4\pi R dR/R]\times (\sigma_{\rm T}/
m_e c^3)$.

The {\sc eqpair} code returns both transmitted (unscattered) seed
photons and Comptonised photons together, but the transmitted seed
photons can be separately calculated with a second call to {\sc
  eqpair} using the sign of the seed photon compactness as a
switch$^2$. We use this to separate out the Comptonised and
unscattered seed photons, at each radius, and then add together the
Comptonised seed photons from all anuuli to form a single Comptonised
spectrum for plotting purposes, and for convolution with the {\sc
  rfxconv} and {\sc kdblur} models to produce the self-consistent
ionised reflected emission from this region. Similarly we add together
all the unscattered seed photons from all radii in the coupled
disc-corona region, again so that we can show these as a single
component for plotting.

We also consider the case where the innermost radius of the coupled
disc corona is truncated before the ISCO, where the remaining power is
dissipated in a hot inner flow.  This geometry allows a smooth
transition between the truncated disc models used for the low/hard
state and the VHS geometry. This inner flow region is also required by the
Lense-Thirring precession models for the low-frequency QPO as a
vertical oscillation is suppressed if there is a disc in the midplane
(Ingram, Done \& Fragile 2009).  We again use {\sc eqpair} to model
this Comptonised emission so that this can also be non-thermal/hybrid, 
and set $\ell_{\rm h}$ from the remaining accretion power between where the disc truncates,
$R_{\rm flow}$, and $R_{\rm ISCO}$ i.e. we assume 
$L_{\rm h}=f_{\rm rad}\int_{R_{\rm flow}}^{R_{\rm ISCO}}L_{\rm NT}(R) 4\pi R dR$ where $f_{\rm rad}\leq 1$ is a factor to allow the
flow to be radiatively inefficient as might be the case if it powers a
strong jet. This sets $\ell_{\rm h}=L_{\rm h}/R_{\rm flow} \times \sigma_{\rm T}/(m_ec^3)$.

In this inner region, we only calculate a single {\sc eqpair} model as 
we assume a single seed photon temperature which is set by that of the 
innermost edge of the disc  i.e. that they are a blackbody with temperature equal to that
of the disc in the coupled disc-corona region (or the standard outer
disc if $R_{\rm cor}=R_{\rm flow}$ when the coupled coronal region does not
exist). These seed photons 
for the inner region are produced in at different
radius, so their energy adds to that released in the inner region. It
is not removed from the region where the seed photons are generated as
these photons are intercepted by the corona along a different line of
sight than is observed. Hence only the Comptonised photons are seen,
as any seed photons which are transmitted rather than scattered are
not in the observers direction. The fraction of seed photons which
intercept the flow depends on the detailed structure of the corona
(scale height, and radial electron emissivity). Since these are
unknown, we set $\ell_{\rm h}$ as above, then have $\ell_{\rm h}/\ell_{\rm s}$ as a free
parameter and then determine after the fit whether the required $L_{\rm s}$
is physically reasonable e.g. a spherical source within a truncated
disc can intercept at most $1/3$ of the photons from the disc\citep{gardner13}. 
Again we separate out the Comptonised photons (using
the sign of $R_{\rm flow}$ as a switch) so as to be able to compute the
self-consistent reflected emission.

\begin{figure}
\includegraphics[width=80mm]{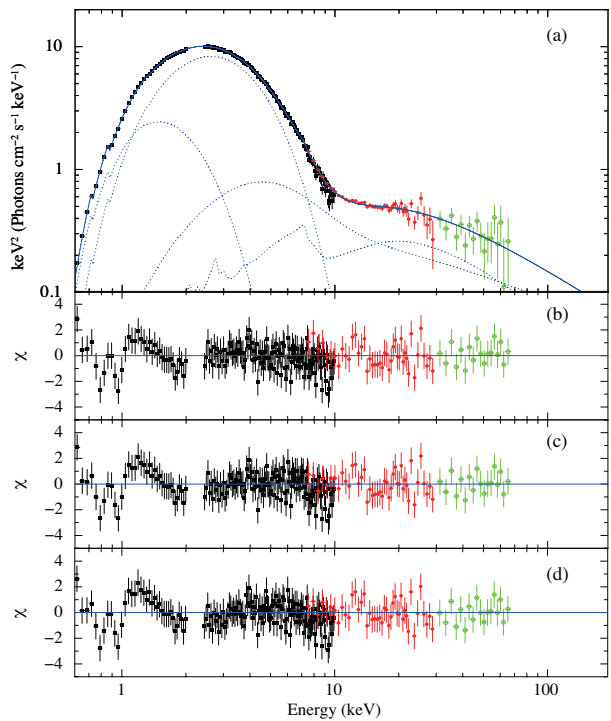}
 \caption{$\nu F_\nu$ spectrum and residuals between the data and the best fit models of GX$339-4$ in the high/soft state.  Data points of pn, PCA, and HEXTE, are indicated with black filled square, red filled circle, and green open circle, respectively. 
 The $\nu F_\nu$ spectrum is based in the best fit {\sc diskEQ} with hybrid corona (panel a).  
 Residuals between data and the best fit {\sc kerrbb}, relativistically smeared {\sc optxeq} without any corona, 
 and  {\sc optxeq} with corona are shown in panels (b), (c), and (d), respectively. }
\label{fig:xmm}
\end{figure}

\section{Calibrating the emissivity from the high/soft state}

To investigate geometry of the disc and corona in the VHS, the
standard value of the spin (or an value of the ISCO) is required.  In
this section, we present the analysis of the high/soft state spectrum
obtained with XMM and RXTE.

\begin{table*}
\begin{minipage}{175mm}
 \caption{The best fit parameters for the high soft state spectrum fitted with {\sc simpl*kerrbb} (column 1), {\sc simpl*diskbb} (column~2), relativistically smeared {\sc simpl*diskEQ} without coupled corona (column~3), and {\sc diskEQ} with coupled corona (column~4).}
 \label{tab:hs}
 \begin{tabular}{@{}llllllll}
 \hline\hline
&  &(1)  &(2)  &(3)  & (4)\\  
component & parameter & {\sc simpl*kerrbb}$^a$& {\sc simpl*diskbb}  &{\sc simpl*diskEQ$^d$ } &\sc{diskEQ(disc-corona)}\\ 
\hline
{\sc tbabs} & $N_{\rm H}(10^{21}~{\rm cm^{-2}})$ &$6.86^{+0.04}_{-0.03}$ &$6.57\pm0.05$   &$6.77\pm0.05$&$6.72^{+0.05}_{-0.02}$ \\
{\sc simpl} &$\Gamma$ &$2.64^{+0.05}_{-0.09}$&$2.61^{+0.10}_{-0.09}$ &$2.55^{+0.10}_{-0.09}$&-----\\
	&$f(10^{-2})$&$3.2^{+0.2}_{-0.1}$ &$3.3\pm0.2$  &$3.0\pm0.2$&-----\\
{\sc rfxconv}$^a$ & $\Omega/2\pi$&$1.3^{+0.5}_{-0.4}$&$1.1^{+0.5}_{-0.3}$ &$1.0^{+0.4}_{-0.3}$&$1.7^{+0.5}_{-0.2}$\\
&$\log \xi$&$<1.72$&$2.61^{+0.11}_{-0.14} $ &$2.35^{+0.03}_{-0.04}$&$2.0^{+0.2}_{-0.4}$\\
{\sc diskbb}  &$T_{\rm in}$~(keV)&-----&$0.868^{+0.003}_{-0.002}$&-----&-----\\
&$r_{\rm in}$~(km)&----- &$54.2\pm0.4$&-----&-----\\
{\sc kerrbb}$^b$  &$l_{\rm h}/l_{\rm s}$    &-----&-----&-----&$0.068^{+0.003}_{-0.004}$\\
~~or {\sc diskEQ}$^c$ &$\tau_{\rm cor}$    &-----&-----&-----&$0.13^{+0.03}_{-0.02}$\\
&$R_{\rm cor}(R_{\rm G})$    &-----&-----&-----&$46^{+9}_{-8}$\\
&$\dot{M}( 10^{18}~{\rm g~s^{-1}})$&$4.33\pm0.04$ &-----&$3.93^{+0.19}_{-0.18}$&$3.99^{+0.05}_{-0.02}$\\
&$a^*$&$0.115\pm0.010$&-----&$0.057^{+0.017}_{-0.015}$&$0.066^{+0.033}_{-0.035}$\\
  \hline
   $\chi^2/dof$ &&$205.2/213$&288.4/213  &208.7/213 &211.3/212\\
 \hline	
 absorbed 0.7--100keV flux&${\rm 10^{-8}~erg~s^{-1}cm^{-2}}$&2.36&2.36&2.36&2.36\\
 unabsorbed 0.7--100keVflux&${\rm 10^{-8}~erg~s^{-1}cm^{-2}}$&3.27&3.21&3.26&3.24\\
 unabsorbed 0.1-200 flux&${\rm 10^{-8}~erg~s^{-1}cm^{-2}}$&4.11&3.93&4.01&3.99\\
\hline
\end{tabular}

 \medskip
 Note: 
 Energy is extended from 0.1~keV to 1000keV. Distance and inclination are assumed to be $D=8$~kpc and $i=50^\circ$.  
 $^a$The parameters of the smeared reflection
are fixed at iron abundance $A_{\rm Fe}=1$, inclination $i=50^\circ$,
$R_{\rm in}=6R_g$, $R_{\rm out}=400R_g$, and power-law illumination
emissivity index of $\beta=3$.
 $^b$ While returning radiation is not included,  limb-darkening is included. 
$^c$ The hot inner flow is not included. A normalization factor, $\ell_{\rm nth}/\ell_{\rm h}$, and $\Gamma_{\rm inj}$ are fixed at 1.287, 1, and 3.5, respectively.
$^d$ {\sc diskEQ} spectrum is relativistically smeared by convolved with {\sc kerrconv}, in which all the parameters are fixed (i.e., $\beta_1=\beta_2=-3$, break radius as $6R_g$, $r_{\rm in}=R_{\rm ISCO}$ and $r_{\rm out}=2R_{\rm ISCO}$).
 \label{tab:xmm}
 \end{minipage}
\end{table*}


\begin{figure*}
\includegraphics[width=58mm]{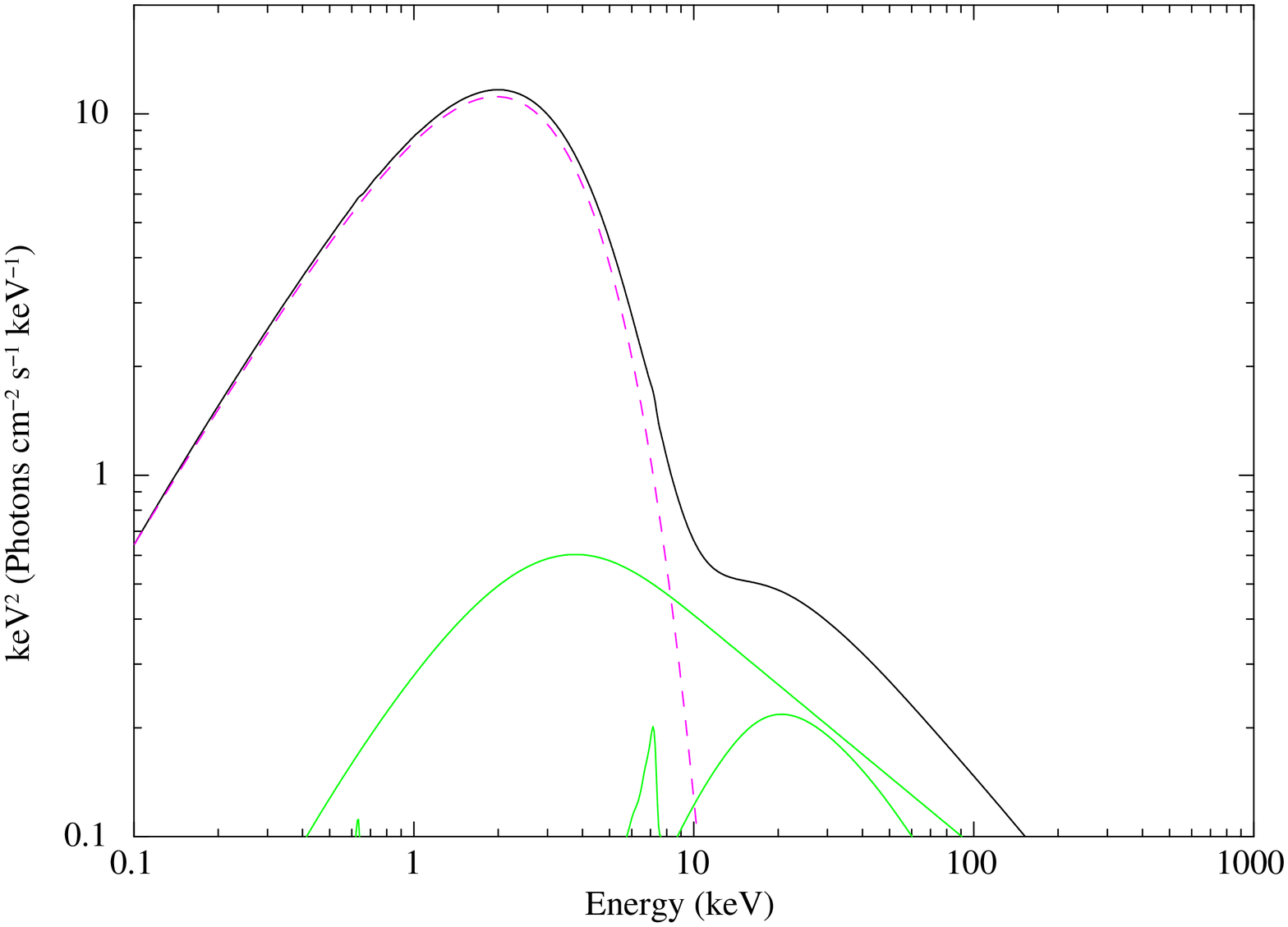}t
\includegraphics[width=58mm]{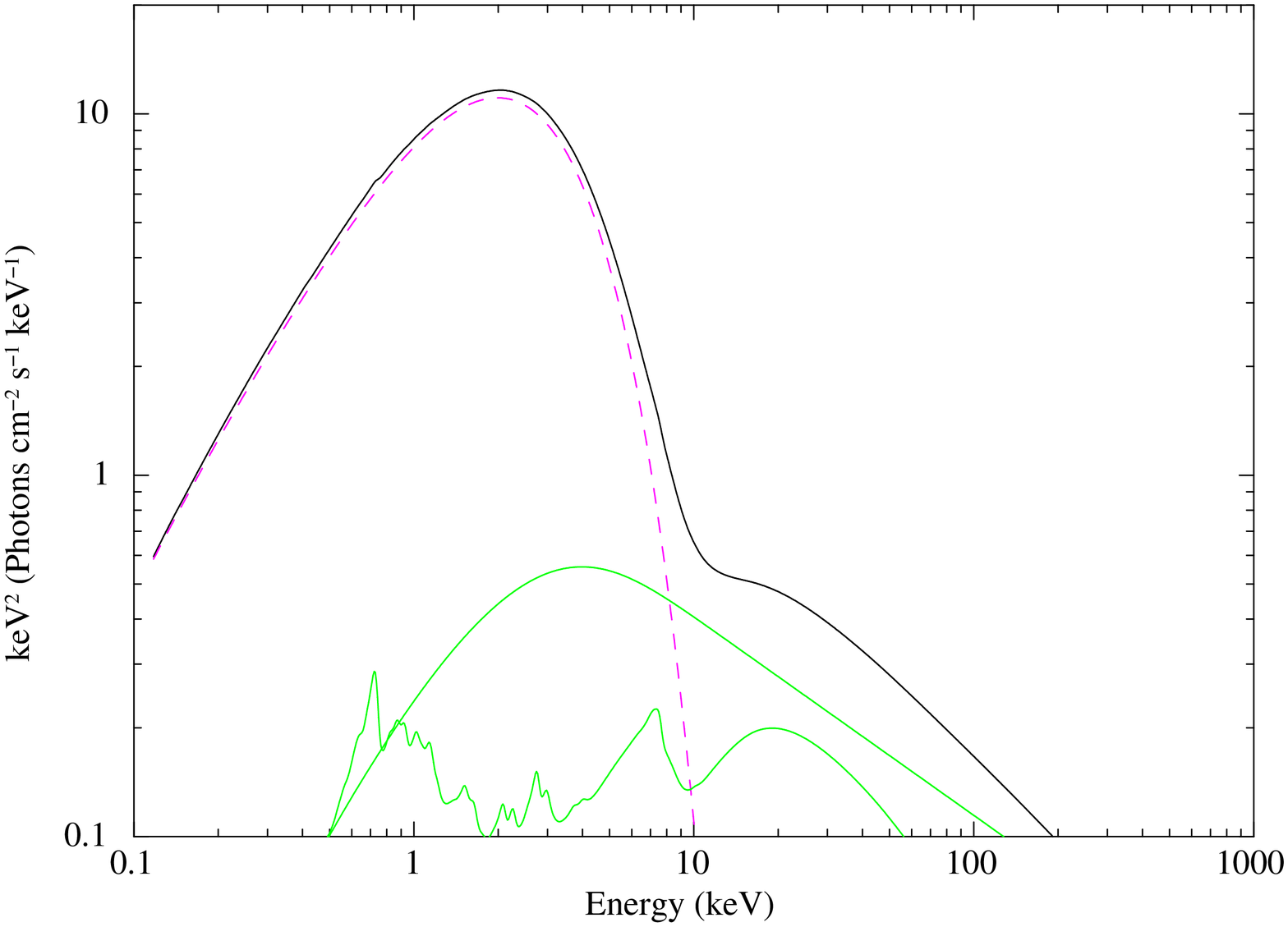}
\includegraphics[width=58mm]{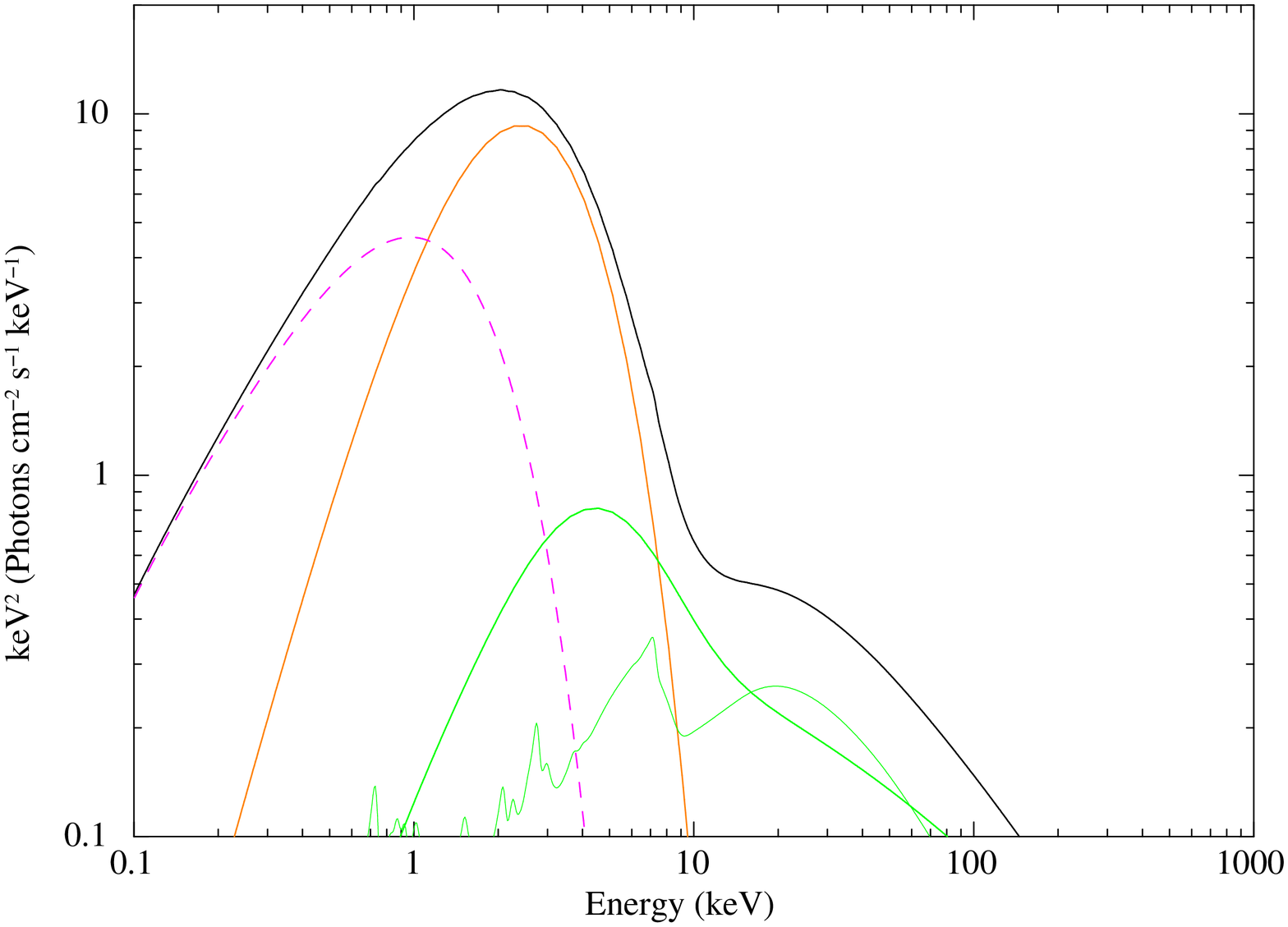}
 \caption{Unabsorbed spectral models for the high/soft state spectrum
with XMM and RXTE, based on best fit models of {\sc kerrbb} with
  {\sc simpl} comptonisation (left),
relativistically smeared {\sc diskEQ} with inner corona (middle),
and {\sc diskEQ} with both coupled and inner corona (right). 
The model components shown separately are the disc (dashed magenta),
inner coronal (or {\sc simpl}) emission + its reflection (both solid
green), and the coupled corona (orange, right panel only).}
\label{fig:xmm_dummy}
\end{figure*}

\subsection{Standard Modeling}

In order to characterize the spin parameter based on the standard
modeling of emission from the optically thick and geometrically thin
disc, we first model the disc emission in the high/soft state with a
Novikov-Thorne emissivity disc {\sc kerrbb} \citep{kerrbb}, convolved
with {\sc simpl} \citep{steiner09} to model the weak Compton tail.
Following T12, the distance, mass and inclination are assumed
to be $D=8$~kpc, $M=7M_\odot$ and $i=50^\circ$, respectively.  The
Compton tail illuminates the disc and forms an ionized reflection
component, modelled using the convolution model {\sc rfxconv}, which
calculates the reflected spectrum and iron line from an X-ray
illuminated slab including Compton scattering \citep{mari11}.  The
reflected spectrum was smeared by general relativistic effects using
{\sc kdblur}\citep{laor91}. The parameters of the smeared reflection
are fixed at iron abundance $A_{Fe}=1$, inclination $i=50^\circ$,
$R_{\rm in}=6R_g$, $R_{\rm out}=400R_g$, and power-law illumination
emissivity index of $\beta=3$.

Here and hereafter, all the model components are absorbed by a common
hydrogen column models with the Tuebingen-Boulder ISM absorption model
({\sc tbabs}) under the abundance by \citet{wilms00}.  The model is
described as {\sc tbabs(simpl*kerrbb+ kdblur*rfxconv*simpl$_{\rm
comp}$* kerrbb)} in {\sc xspec}.  To use the convolution models, {\sc
simpl}, {\sc rfxconv}, and {\sc kdblur}, we extended the energy band
from 0.1~keV to 1000~keV.
The {\sc kerrbb} model with limb-darkening and without returning
radiation reproduced the spectrum well with $\chi^2/dof=205.2/213$,
and gave an absorbed 0.7--100~keV flux of $2.36\times 10^{-8}~{\rm erg~s^{-1}cm^{-2}}$. 
All the parameters are given in column (1) in Table~\ref{tab:xmm}, 
and the best fit spin parameter of $a^\ast=0.11^{+0.02}_{-0.01}$ indicates $R_{\rm ISCO}$ of $5.6R_g$. 
The residuals between the data and
model are shown in Figure~\ref{fig:xmm}(b), and unabsorbed model
spectrum is shown in left panel of Figure~\ref{fig:xmm_dummy}.

We replaced the {\sc kerrbb} model with the simpler {\sc diskbb}. This
did not give an acceptable fit, with $\chi^2/dof=288.4/213$ (see column (2) in Table~\ref{tab:xmm}), as the
relativistic effects included in {\sc kerrbb} give a significantly
broader disc spectrum. However the derived apparent inner radius of
$r_{\rm in}=54$~km is consistent with the ISCO of non-spinning black
hole, $6R_{\rm g}$, of $7M_\odot$.  Here the true inner radius is
estimated by $\kappa ^2\xi r_{\rm in}$ as $\sim 63$~km with the color
hardening factor $\kappa=1.7$ \citep{shimura95} and the correction
factor for inner boundary condition of $\xi=0.41$ \citep{kubota98}.  We
caution that the absolute values of spin depend strongly on the assumed
black hole mass, distance, and inclination (see e.g. Kolehmainen et al. 2010)
but here we are interested only in comparative rather than
absolute radii as the focus of this paper is whether inner radius of
the thin disc component is truncated in the VHS with respect to the
disc dominated high/soft state.

\subsection{{\sc diskEQ} without the corona}

We replaced the {\sc kerrbb} model in the fits above with our new disc
corona model {\sc diskEQ}. First we use this without any coronal
emission, i.e. using this a purely Novikov-Thorne optically thick disc
like {\sc kerrbb}, but without the relativistic corrections.  {\sc diskEQ}, similarly to {\sc optxagnf} \citet{optxagn}, is for an angle
averaged disc, so we incorporate inclination by fixing the
normalisation to $\cos i/0.5=1.287$ rather than unity.  We use this together
with the {\sc simpl} and {\sc rfxconv} models to describe the weak
Compton tail and its reflection.  We get a poor fit, with
$\chi^2/dof=297.0/213$, similar to that derived from the {\sc diskbb}
fit above, but again the spin is low, at $a^*=0.12$.  This shows
firstly that the self-consistent colour temperature correction used in
{\sc diskEQ} is close to 1.7 as assumed for the {\sc kerrbb} fits, and
secondly that it is indeed relativistic smearing as opposed to the
detailed shape of the Novikov-Thorne emissivity 
which gives the better fit of {\sc kerrbb} compared to {\sc diskbb}. 

We can approximate the effect of relativistic smearing into the disc
model using {\sc kerrconv} \citep{kerrconv} by tying its spin
parameter to that of {\sc diskEQ}, fixing the emissivity index to
$-3$, inner convolution radius equal to $R_{\rm ISCO}$ and the outer at
$2R_{\rm ISCO}$ \citep{done13}. This improves the fit significantly
with $\chi^2/dof=208.7/213$ (Table~\ref{tab:xmm}), but the best fit value of the spin is
similar, at $a^*=0.06$.  Thus relativistic corrections can affect the
quality of the fit, but do not significantly change the best fit
parameters. The  residuals between the data and this
model are shown in Figure~\ref{fig:xmm}(c), the unabsorbed model
spectrum 
is shown in middle panel of Figure~\ref{fig:xmm_dummy} and all
parameters are detailed in column (3) in Table~\ref{tab:xmm}.

\subsection{{\sc diskEQ} with the corona}

We now remove the {\sc simpl} description of the Compton tail, and
instead describe it self consistently using the {\sc diskEQ} model,
with $R_{\rm in}=R_{\rm ISCO}$ (see schematic geometry in the left hand
panel of Figure~\ref{fig:nt}). We reflect only the coronal emission
via the  switch in the code, using the convolution models
{\sc rfxconv} and {\sc kdblur} as before.  The model is described as
{\sc TBabs(diskEQ + kdblur*rfxconv*diskEQ$_{\rm comp}$)} in {\sc xspec}.

To replicate the shape of {\sc simpl} under the description of the
{\sc diskEQ}, we fix $\ell_{\rm nth}/\ell_{\rm h}=1$, i.e. assume completely
non-thermal electron injection so as to get closest to the completely
power law Compton tail assumed in {\sc simpl}. The {\sc simpl} code
also assumes that the entire disc emission at all radii is equally
Compton scattered, so we set $R_{\rm cor}=500R_g$. We fix
$\Gamma_{\rm inj}=3.5$ (as required from a similar VHS spectrum from
XTEJ1550-564; Hjalmarsdotter et al. 2016) as this is not well
constrained by our data. This gives $\chi^2/dof=246.3/213$ as the
Comptonisation in {\sc eqpair} has a hybrid rather than pure power law
shape even for completely non-thermal injection. Coulomb collisions
act to thermalise the electrons at low energy, giving rise to a low
temperature thermal Compton component which effectively smooths the
transition between the Wien disc spectrum and power law tail. 

The fit improves still further to $\chi^2/dof=211.3/212$ by letting $R_{\rm cor}$ free.  The unabsorbed model components are shown in the right
panel of Figure~\ref{fig:xmm_dummy}. In addition to the low
temperature thermal Comptonisation from the hybrid electron
distribution (green), the best fit $R_{\rm cor}\sim 46R_g$ means that
the outer disc (magenta) is seen directly, while only the inner disc
under the corona (orange) is slightly suppressed by the weak Comptonisation, again
resulting in a slight broadening of the observed disc emission.  This
fit is shown as the baseline $\nu F_\nu$ spectrum in
Figure~\ref{fig:xmm}(a), while its residuals are shown in
Figure~\ref{fig:xmm}(d).  The best fit spin is again
$a^\ast=0.07$, and all parameters are given in column (4) in Table~\ref{tab:xmm}.

In conclusion, at these values of spin and inclination, the lack of relativistic
effects on the disc emission is not as significant as the detailed
geometry assumed for the disc and corona. Hence in all that follows we
use {\sc diskEQ} without relativistic smearing. However, we also note
that the derived reflection parameters are  dependent on the
details of how the continuum spectra are modeled e.g. the reflection
solid angle is $\Omega/2\pi \sim 1$ with relativistic smearing of the
disc continuum, significantly lower than the value ($\Omega/2\pi\sim
1.7$) derived without including this.  This is because the
relativisitic smearing slightly broadens the disc emission, so the
disc makes more of the continuum in the 5--8~keV range,  requiring
less reflection.

\section{The very high state}

\subsection{Overview}

\begin{figure}
\includegraphics[width=41mm]{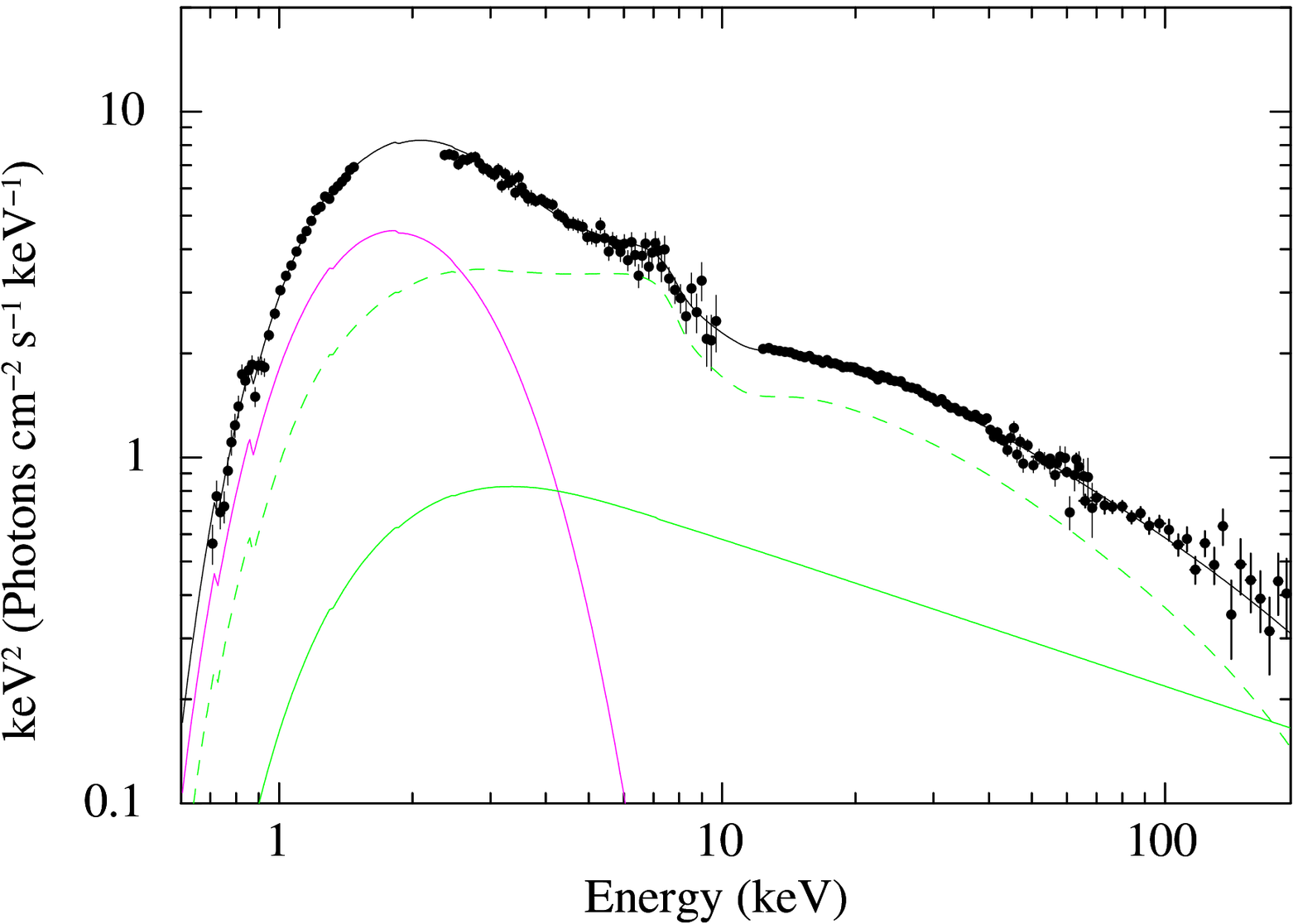}
\includegraphics[width=41mm]{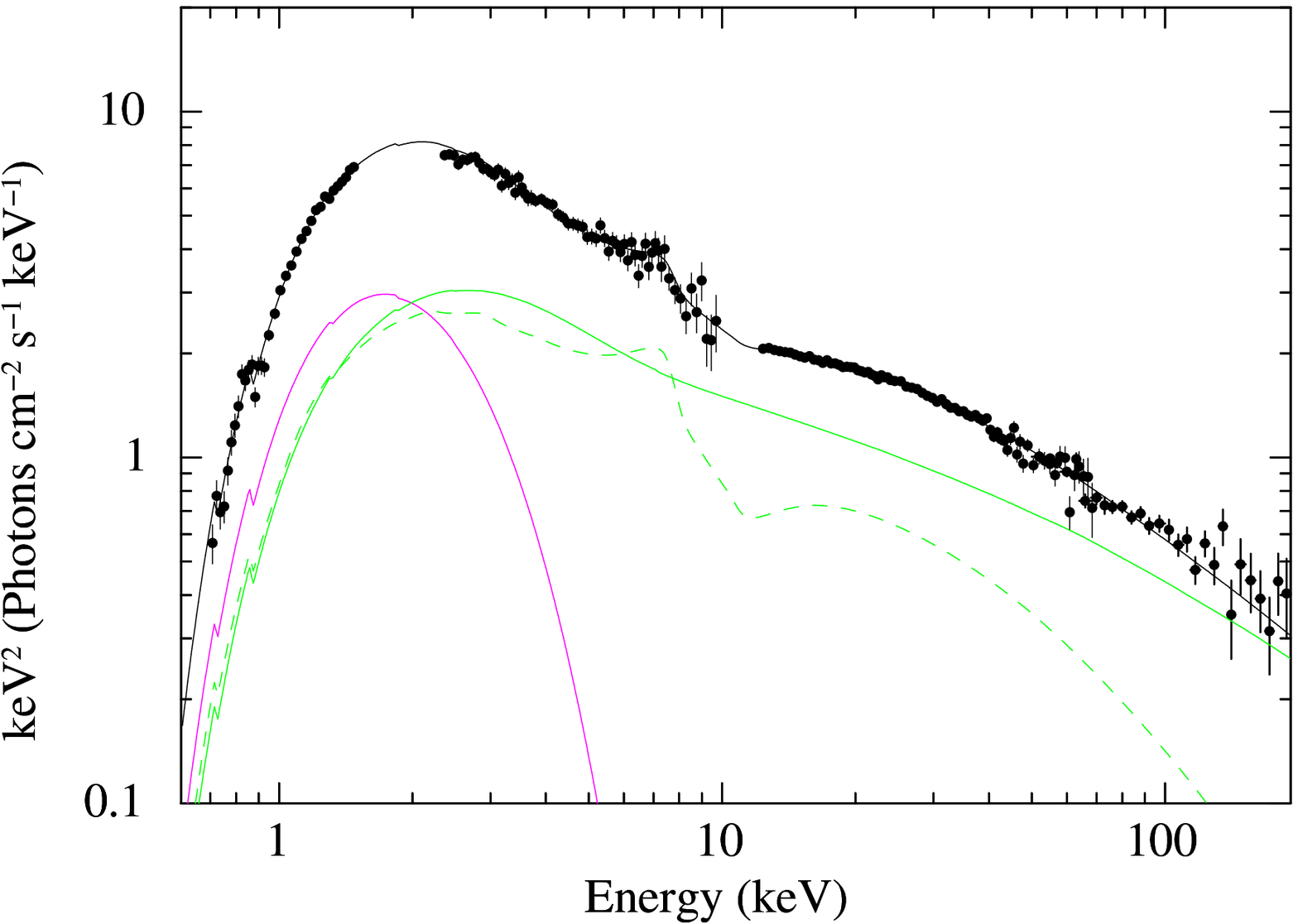}
 \caption{The Suzaku VHS spectrum fitted with {\sc simpl*diskbb}
 (left) and {\sc diskbb+eqpair} (right). The components are
{\sc diskbb} (solid magenta), comptonised
 {\sc diskbb} or compton component of {\sc eqpair} (solid green), and reflected
 emission (dashed green). } 

 \label{fig:overview}
\end{figure}

We first illustrate the data using the {\sc simpl} description of
Comptonisation, with 
{\sc tbabs(simpl*diskbb + kdblur*rfxconv * simpl$_{\rm comp}$*diskbb)}.  
This assumes that the seed photons are the
entire disc, so all temperatures are equally scattered. 
This is an acceptable fit to the data, 
with parameters which require that the fit is reflection dominated,
with $\Omega/2\pi=4.8$, and that this reflection is highly smeared,
with $R_{\rm in}=1.5$ and $\beta=3.5$. This is in agreement with the study
of Ludlam et al (2015), where they use this model to argue that these
data require a high spin black hole. The spectral decomposition using
this model is shown in a left panel of Figure~\ref{fig:overview}, and details
of the fit  are given in Table~\ref{tab:overview}.  

We then replace the {\sc simpl} model with the {\sc eqpair} description of
Comptonisation, where the spectrum is not a pure power law even for
completely non-thermal acceleration. 
We assume that the seed photons
are blackbody at the maximum disc temperature $T_{\rm in}$.  
This gives a better
fit to the data, and completely different reflection parameters. The
amount of reflection is no longer enhanced, with $\Omega/2\pi =0.8$, and
there is little relativistic smearing. This illustrates how the spin
parameter derived from reflection fits can be strongly affected by the
choice of continuum. There is complex curvature in the data, which can
be fit either by a large amount of highly smeared reflection on a
power law Comptonisation continuum or by less reflection on the more
physical {\sc eqpair} continuum. We show this very different
decomposition in a right panel of Figure~\ref{fig:overview} with details of all parameters in  Table~\ref{tab:overview}.
We caution that the reflection parameters are not always robust to a
different continuum choice.

\begin{figure}
\includegraphics[width=80mm]{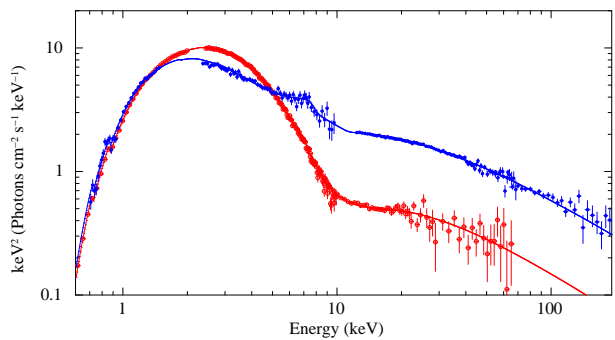}
 \caption{Comparison of $\nu F_\nu$ spectra of the high/soft state (red) and the VHS (blue). This is basically same as figure~7 in T12, but based on the best fit models of {\sc diskEQ} with coupled corona and {\sc diskbb+eqpair}, respectively. }
 \label{fig:comparison}
\end{figure}

What is constant between both fits is the very low disc temperature,
at $\sim $0.5--0.6~keV.  Figure~\ref{fig:comparison} shows comparison
between the high/soft state spectrum and the VHS spectrum based on the
best fit models of {\sc diskEQ} with hybrid corona and {\sc
diskbb+eqpair}, respectively.  A disk peaks in $\nu f_\nu$ at
$2.35~kT_{\rm in}$ with i.e. at 1.2--1.4~keV. As shown in
Table~\ref{tab:overview}, the bolometric flux in the VHS is a factor
$1.1\times$ larger than in the high/soft state, so this predicts that
$\dot{M}$ also increases by a factor of 1.1.  Thus a pure disk
spectrum should peak at a very similar (slightly higher)
energy than that 
of the high/soft state.  Instead, as shown in Figure~\ref{fig:comparison}, the
VHS spectrum peaks at lower energy than the high/soft state.
It is this decreasing energy at which the spectrum peaks
which causes the models to struggle to fit these data. 

Simply overlaying the inner disc with the Comptonising corona does not
help. Steep spectrum Comptonisation retains the imprint of the seed
photons, with a low energy turnover at $7kT_{\rm bb}$ for
$\Gamma=2.5$. 
On the contrary, the observed spectral peak is determined by the
sum of this and the disc, with equal energy in both leading to a
spectral peak at $\sim 2$~keV for a peak disc temperature of 0.53~keV
(see the right panel of Figure~\ref{fig:overview}).
Thus the seed photons for the Comptonisation must be no more than
$\sim 0.5$~keV, when the expected inner disc temperature is $\sim
1$~keV. Covering the inner disc and using its energy to power
the corona seems to be the obvious solution.  However, the decrease in
temperature produced by using a fraction $g$ of the locally released
gravitational energy to power the corona is $T=(1-g)^{1/4}T_0$, so
requires $g=0.87$. This means that the majority of the power is
released in the coronal region rather than in the underlying disc,
though the steep spectrum shows very clearly that there is more energy in the
seed photons than in the Comptonised emission. 

A steep spectrum with
$\Gamma\sim 2.5$ requires $\ell_{\rm h}/\ell_{\rm s}\sim 0.5$,
i.e. $g=\ell_{\rm h}/(\ell_{\rm s}+\ell_{\rm h})\sim 0.3$ (T12) whereas $g=0.87$ implies
$\ell_{\rm h}/\ell_{\rm s}\sim 7$. The soft Comptonised spectral shape is then at
odds with the strong energy dissipation in the corona required for the
underlying disc temperature to drop sufficiently to match the data. 
We note that 
including an albedo less than unity (which would be required for a spectrum as hard as $\ell_{\rm h}/\ell_{\rm s}=7$)
will only strengthen this conclusion, as
the thermalised emission adds to the intrinsic seed photons to steepen the spectrum. 

\begin{table*}
\begin{minipage}{120mm}
 \caption{Overview of the Suzaku VHS spectrum}
 \label{tab:overview}
 \begin{tabular}{@{}llllllll}
 \hline\hline
component & parameter & {\sc simpl*diskbb}  & {\sc diskbb+eqpair}\\
\hline
{\sc tbabs} & $N_{\rm H}(10^{21}~{\rm cm^{-2}})$ &$6.72^{+0.19}_{-0.12}$&$6.91^{+0.15}_{-0.31}$\\
{\sc kdblur}&$\beta$&$3.5\pm0.4$&(3.0)\\
&$R_{\rm in} (R_g)$&$1.5^{+0.9}_{-0.3{\rm (hard~limit)}}$&$4.9^{+3.3}_{-1.4}$\\
{\sc rfxconv} & $\Omega/2\pi$&$4.8\pm1.1$&$0.82^{+0.09}_{-0.12}$\\
&$\log \xi$&$3.71^{+0.07}_{-0.09}$&$4.0^{+0 {\rm (hard~limit)}}_{-0.1}$\\
{\sc diskbb}  &$T_{\rm in}$~(keV)&$0.580^{+0.013}_{-0.012}$&$0.526_{-0.008}^{+0.014}$\\
&$r_{\rm in}$~(km)&$94\pm5$&$92^{+4}_{-7}$ \\
{\sc simpl} &$\Gamma$ &$2.427^{+0.017}_{-0.021}$&-----\\
	&$f(10^{-2})$&$7.8^{+2.1}_{-1.8}$&-----\\
{\sc eqpair}$^a$ &$\ell_{\rm h}/\ell_{\rm s}$    &-----&$0.28\pm0.02$\\
&$\tau_{\rm}$    &-----&$0.67^{+0.08}_{-0.05}$\\
&norm&-----&$0.33^{+0.11}_{-0.03}$\\
\hline
  $\chi^2/dof$& &228.9/198&214.5/198\\
 \hline	
 absorbed 0.7--100~keV flux &$10^{-8}~{\rm erg~s^{-1}~cm^{-2}}$&2.65&2.64\\
 unabsorbed 0.7--100~keV flux &$10^{-8}~{\rm erg~s^{-1}~cm^{-2}}$&3.58&3.61\\
 unabsorbed 0.1--200~keV flux &$10^{-8}~{\rm erg~s^{-1}~cm^{-2}}$&4.55&4.72\\ 
 \hline\hline
\end{tabular}

 Note: Values in parenthesis are fixed.  
$^a$ Values of $\ell_{\rm nth}/\ell_{\rm h}$, $\gamma_{\rm min}$, $\gamma_{\rm max}$, and 
$\Gamma_{\rm inj}$ are fixed at 1, 1.3, 1000, and 3.5, respectively.  
\end{minipage}
\end{table*}

\begin{table*}
\begin{minipage}{175mm}
 \caption{Same as Table~\ref{tab:xmm} but for the Suzaku VHS spectrum with {\sc diskEQ}.  }
 \label{tab:vhs}
 \begin{tabular}{@{}llllllll}
 \hline
 \hline
 &  & \multicolumn{2}{c}{ outer disk + disc-corona } &   \multicolumn{2}{c}{outer disc+ disc-corona + flow}\\ 
\cline{3-6}
component & parameter &(1) free spin &(2) scale correction &(3) untruncated &(4) truncated\\
\hline
{\sc tbabs}  &$N_{\rm H}(10^{21}~{\rm cm^{-2}})$ &$6.85^{+0.14}_{-0.15}$&$6.93^{+0.22}_{-0.15}$&(6.8)&(6.8)\\
{\sc rfxconv}$^a$ & $\Omega/2\pi$&$1.59^{+0.19}_{-0.17}$&$1.61^{+0.18}_{-0.17}$&$0.61^{+0.16}_{-0.13}$&$0.68^{+0.09}_{-0.14}$\\ 
{\sc diskEQ}$^b$&$l_{\rm h}/l_{\rm s}$    &$0.45\pm0.02$&$0.44^{+0.03}_{-0.05}$&$0.46^{+0.03}_{-0.10}$&$0.45^{+0.02}_{-0.05}$\\
&$\tau_{\rm cor}$    &$0.67\pm0.04$&$0.63^{+0.07}_{-0.13}$&$0.46^{+0.09}_{-0.23}$&$0.46^{+0.05}_{-0.10}$\\
&$\tau_{\rm in}$ &---&---&$2.78^{+0.11}_{-0.25}$&$2.66^{+0.07}_{-0.43}$\\
&$f_{\rm rad}$ &---&---&$0.26\pm0.03$&(1)\\
 &$R_{\rm cor}(R_{\rm g})$    &$48^{+4}_{-5}$&$50^{+7}_{-4}\times scale$&$49^{+8}_{-5}$&$46^{+7}_{-4}\times scale$\\&$R_{\rm flow}(R_g)$ &---&---&$24.5^{+1.8}_{-2.4}$&$18.5^{+1.8}_{-0.7}\times scale$\\
& $\dot{M}~( 10^{18}~{\rm g~s^{-1}})$&$6.48^{+0.11}_{-0.13}$&$6.0^{+0.7}_{-1.1}\times scale$&$5.97^{+0.16}_{-0.25}$&$6.3^{+0.6}_{-0.7}\times scale$\\
&$a^*$    &$-1^{+0.15}_{-0~({\rm hard~limit})}$&($-1$)&(0)&($-1$)\\
&  $scale$&(1)&$1.05^{+0.11}_{-0.07}$&(1)&$1.01^{+0.07}_{-0.05}$\\
\cline{2-6}\\
&estimated $R_{\rm in}(R_g)^c$    &$9^{+0({\rm hard~limit})}_{-0.43}$&$9.5^{+0.9}_{-0.7}$&(6)&$9.0^{+0.7}_{-0.4}$\\
\hline
 absorbed 0.7--100~keV flux&$10^{-8}{\rm erg~s^{-1}cm^{-2}}$&2.62&2.63&2.61&2.61\\
 unabsorbed 0.7--100~keV flux&$10^{-8}{\rm erg~s^{-1}cm^{-2}}$&3.56&3.59&3.54&3.55\\
 unabsorbed 0.1--200~keV flux &$10^{-8}{\rm erg~s^{-1}cm^{-2}}$&4.56&4.61&4.52&4.54\\ 
 \hline
  $\chi^2/dof$ &&228.1/200&224.8/200 &184.0/199&183.6/199\\
 \hline	
 \hline	
 \end{tabular}
 \medskip 
 
 Note: Values in parenthesis are fixed.  Energy is extended from 0.1 keV to 1000~keV. Distance and inclination are assumed to be 
 $D = 8$~kpc and $i=50^\circ$. 
$^a$ Smeared reflection is same as Table~1, but ionization parameter is fixed at $\xi=300$.  {\sc kdblur} is used with $R_{\rm in}=10R_{\rm g}$ and $R_{\rm in}=R_{\rm flow}$ for with and without inner flow, respectively.
 $^b$Values of normalization factor, $\ell_{\rm nth}/\ell_{\rm h}$, 
 and $\Gamma_{\rm inj}$, are fixed to be 1,287, 1, and 3.5, respectively. 
$^c$ The value of $R_{\rm in}$ is estimated by the ISCO for the given spin parameter $a^*$ as $R_{\rm ISCO}\times scale$ .
%
%
%
 
\end{minipage}
\end{table*}

\subsection{{\sc diskeq} with coupled disc-corona over the inner disc }

This tension between the observed spectral peak and observed
luminosity is illustrated when we try to fit the spectrum with a
disc-corona which extends down to $R_{\rm in}=R_{\rm ISCO}$ (left
panel of Figure~\ref{fig:nt}).  We fix the spin to $a_*=0$, as
required in the high/soft state data, and fix the reflection
parameters to $\xi=300$, with smearing of $R_{\rm in}=10R_g$ and $\beta=3$ (all
as used in previous papers including \cite{yamada09} and T12). We
first assume fully non-thermal injection ($\ell_{\rm nth}/\ell_{\rm h}=1$). This
does not give a good fit, with $\chi^2/dof=456.4/201$ for the reasons
outlined above. The steep tail fixes $\ell_{\rm h}/\ell_{\rm s}\sim 0.5$, so
$g\sim 0.3$ and the seed photons for the Comptonisation are only
slightly cooler (factor $0.91$) than for a full disc. Allowing
$\ell_{\rm nth}/\ell_{\rm h}$ to additionally be free did not give a
substantial improvement, with best fit consistent with pure
non-thermal injection. Allowing the disc ionisation to be free switches
the fit to being reflection dominated, which does not then conserve
energy since most of the continuum which produces the observed
enhanced reflection is not included in the energy budget. Since none
of these fits are acceptable, we do not give detailed parameters.

We can have an acceptable fit if we allow the spin to be a free
parameter. The best fit is then $\chi^2=228.1/200$ for $a^\ast=-1$
(i.e., extreme retrograde rotation).  
The detailed parameters are shown in column~(1) in Table~\ref{tab:vhs}. 
Maximally retrograde spin has
$R_{\rm ISCO}=9R_g$, which is 1.5 times larger than the disc dominated
state, possibly indicating that the disc is somewhat truncated in the
VHS or that the stress heating prescription has changed markedly in
the inner disc. However, using spin as a proxy for this transition
radius is somewhat restrictive as this only allows values $<9R_g$. We
allow the model normalisation to additionally be free, and discuss in
the Appendix how to use this to estimate the radius at which the
stresses stop heating the accretion flow. This fits only slightly
better ($\chi^2/dof=224.8/200$), showing that the spectrum is
consistent with a stress free truncation of the emissivity at $\sim
9R_g$. The best fit parameters and the residuals between the data and
model are shown in column~(2) in Table~\ref{tab:vhs} and
Figure~\ref{fig:vhs-diskEQ}(b), respectively. 

\subsection{{\sc diskEQ} with coupled disc-corona and hot inner flow extending down to the ISCO}

The previous section shows that the data are compatible with a model
where the stresses change dramatically in the VHS, so that the flow is
heated with a stress-free boundary at $R\sim 9$--$10R_g$ rather than the
$6R_g$ expected from the known spin calibration. However, this could
also indicate instead that the stresses are as expected for the black
hole spin, but that the accretion flow makes a transition from the
coupled disc-corona to a hot inner flow at radii $>9R_g$, as required
for the QPO models. Also, this would require that this flow is
strongly radiatively inefficient so as to compensate for the
additional energy released between a stress-free truncation at $6R_g$
as expected from Novikov-Thorne, and the  stress-free truncation at
$\sim 9R_g$ which fits the data above. 

We fix $a^*=0$, and $\ell_{\rm nth}/\ell_{\rm h}=1$ but now allow the
disc-corona region to be truncated, with additional inverse Compton
emission from the innermost region between the disc truncation radius
and the ISCO (see the schematic geometry in Figure 1b).  We assume
that the inner region also has completely non-thermal electron
acceleration, with $\ell_{\rm nth}/\ell_{\rm h}=1$, again with fixed
reflection ionisation and smearing (now tied to the inner edge of the
disc-corona, i.e. $R_{\rm flow}$). There are then two remaining parameters
which control the shape of the spectrum from the innner flow, namely
$\tau_{\rm in}$ and $(\ell_{\rm h}/\ell_{\rm s})_{\rm in}$. These cannot be
constrained independently so we choose to tie $(\ell_{\rm h}/\ell_{\rm s})_{\rm in}$
to $\ell_{\rm h}/\ell_{\rm s}$ in the disc-corona region and let $\tau_{\rm in}$ be
free.  In addition, we fixed absorption column at $N_{\rm
H}=6.8\times 10^{21}~{\rm cm^{-2}}$, the average value seen in the high/soft
state. Thus the free parameters increase only by 1. 

This gives an excellent fit to the data (column~(3) in Table~\ref{tab:vhs}), with
$\chi^2/dof=184.0/199$, with reasonable reflection ($\Omega/2\pi=0.61$).
The coupled disc-corona region extends from 49--25$R_g$, with a
radiatively inefficient ($f_{\rm rad}= 0.26$) hot flow from 25--6$R_g$.
The optical depth of the inner corona is estimated at $\tau_{\rm in}=2.8$,
significantly larger than for the disc-corona, where $\tau_{\rm
cor}=0.43$, though we caution that this simply means that the inner
flow has a spectrum which is slightly different to that of the disc
corona. 

The best fit value of $\ell_{\rm h}/\ell_{\rm s}=0.46$ gives 
$g= 0.32$. Together with the best fit $\tau=0.46$ this gives 
an estimate for the intrinsic fraction of power dissipated in the
corona of $f\sim 0.3$--0.4 for an albedo of 1--0
(see Appendix B). 

\subsection{{\sc diskEQ} with coupled disc-corona and efficient hot inner
  flow truncated at the bending wave radius}

Detailed fits to the variability power spectrum of a VHS of similar
QPO frequency in XTE~J1550--564 gives $\sim 13 R_g$ for the transition
between the disc and inner flow (Obs 5 in Ingram \& Done 2012).
However, these QPO models also rely on the misaligned flow not
extending down to the ISCO but being truncated at the bending wave
radius, where the torques become so strong as to truncate the flow.
This is $\sim 12R_g$ in the matching observation of XTE~J1550-564, so
the inner flow in their picture is a narrow ring from $13-12R_g$
(Ingram \& Done 2012). This picture is somewhat different to that
modelled above, where the hot flow which extends down to the ISCO but
can be radiatively inefficient.

It is not at all clear what should happen to the accretion torques in
the inner part of a misaligned flow below the bending wave
radius. There are additional stresses at the bending wave radius, and
then below this the flow free streams into the black hole \citep{fragille07,dexter11}. 
Thus it seems possible that the
stresses might be more like a stress-free inner boundary at the
bending wave radius, or it may be that the additional stresses at the
bending wave radius release all the remaining energy (Fragile, private
communication). The model above assumes the latter, that all the
accretion energy is released in the hot flow, in which case most of
this energy is not radiated as hard X-rays.  Instead, we now explore
the possibility that the flow truncation at the bending wave radius
changes the dissipation to something more like a stress-free inner
boundary condition at the bending wave radius (see Figure 1c). 

\begin{figure}
\includegraphics[width=80mm]{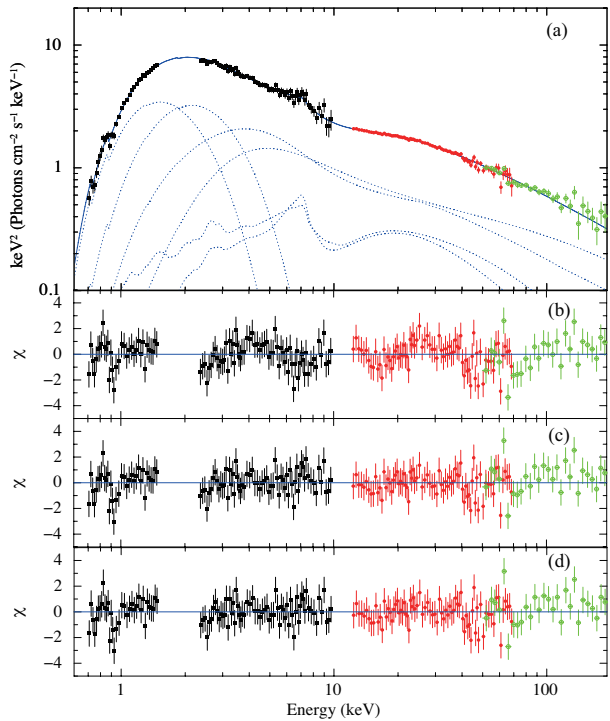}
 \caption{$\nu F_\nu$ spectrum of GX$339-4$ in the VHS based in the best fit {\sc diskEQ}  with stress-free truncated flow (panel a). 
 Residuals between data and the best fit {\sc diskEQ} model without inner flow,  with inefficient inner flow, 
 and with stress-free truncated flow
 are shown in panels (b), (c),  and (d), respectively. 
 In panel (a), PIN (red filled circle) and GSO (green open circle) data are normalized to XIS data (black filled square).}
\label{fig:vhs-diskEQ}
\end{figure}

\begin{figure*}
\includegraphics[width=58mm]{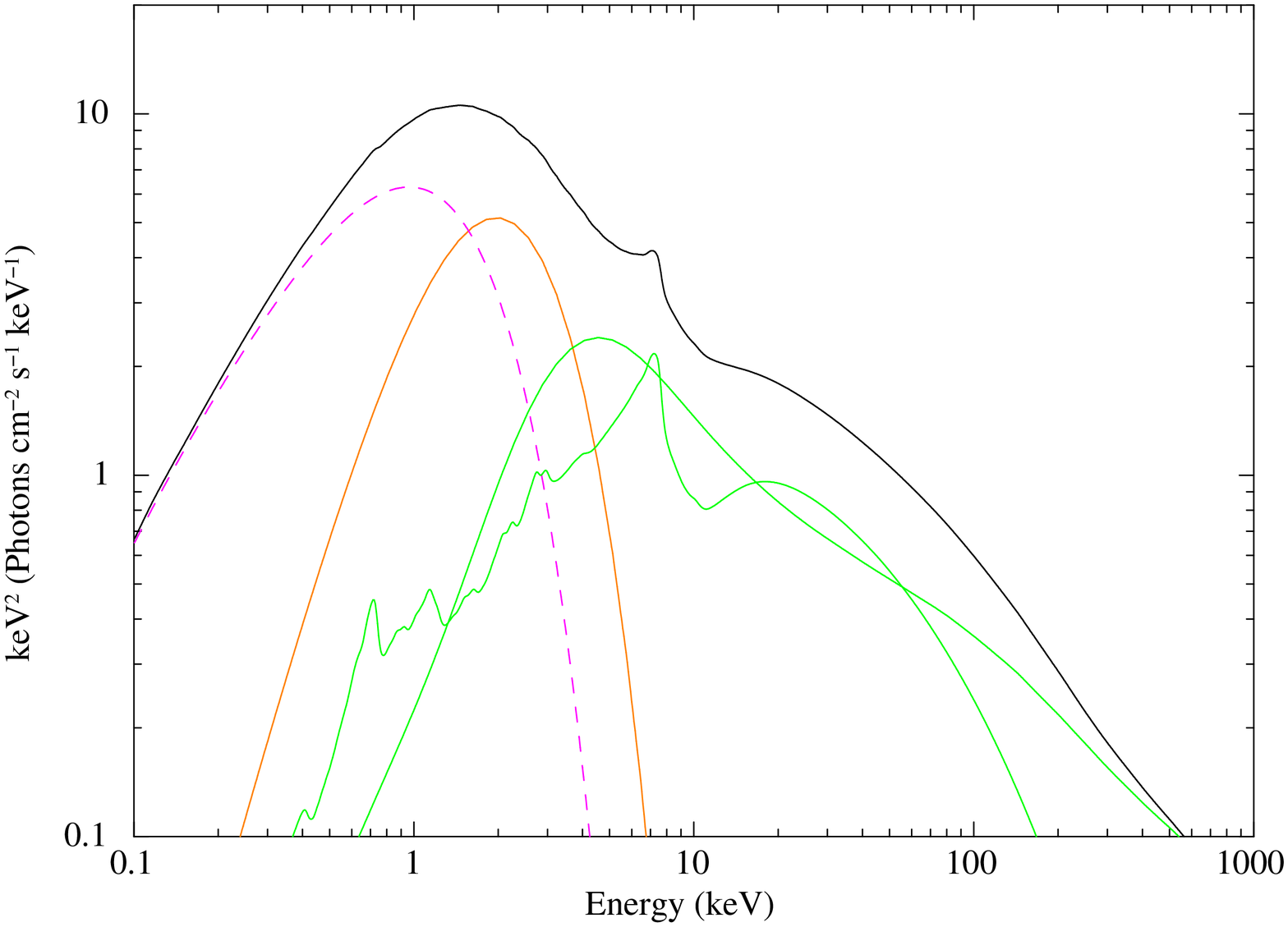}
\includegraphics[width=58mm]{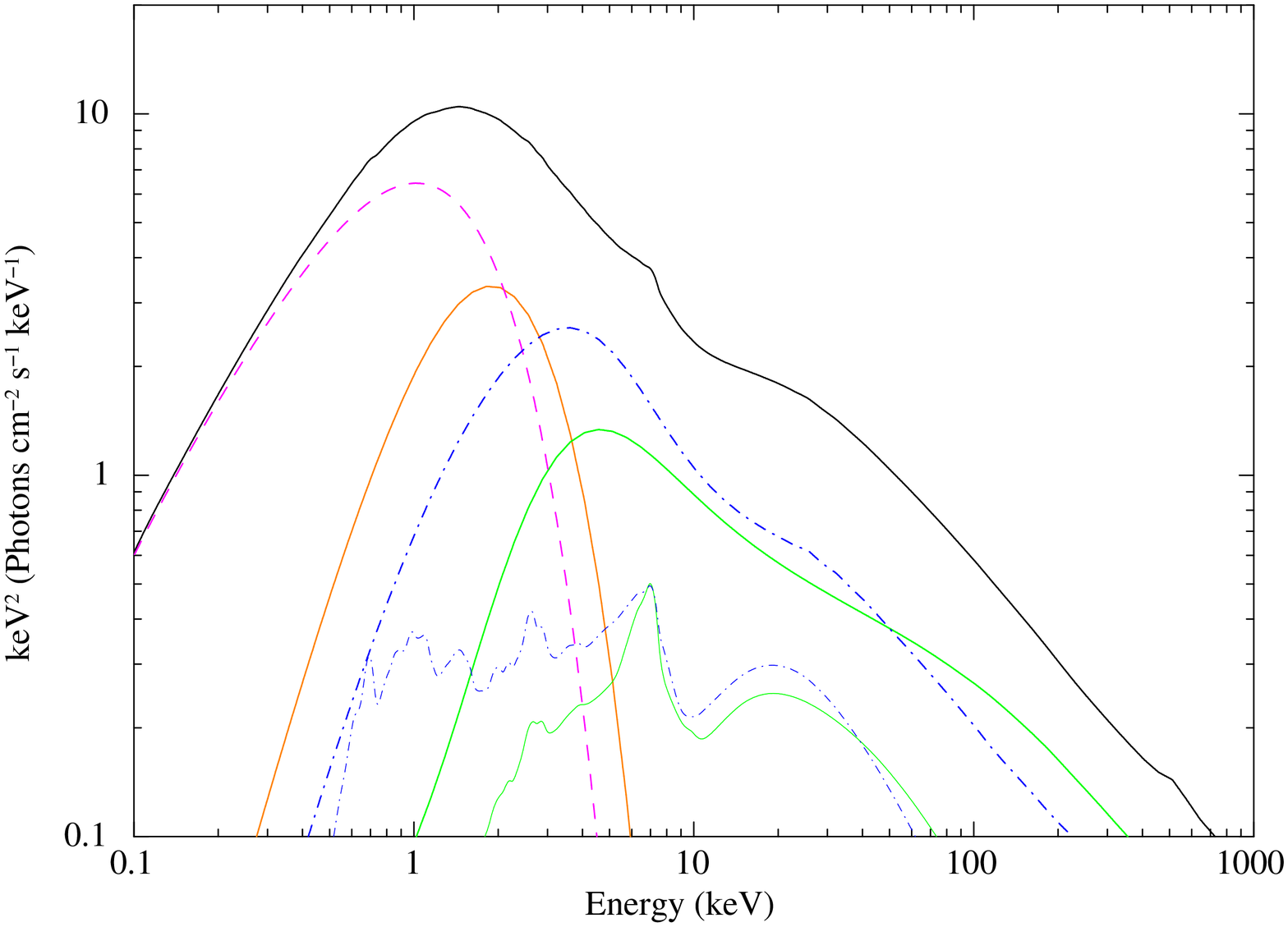}
\includegraphics[width=58mm]{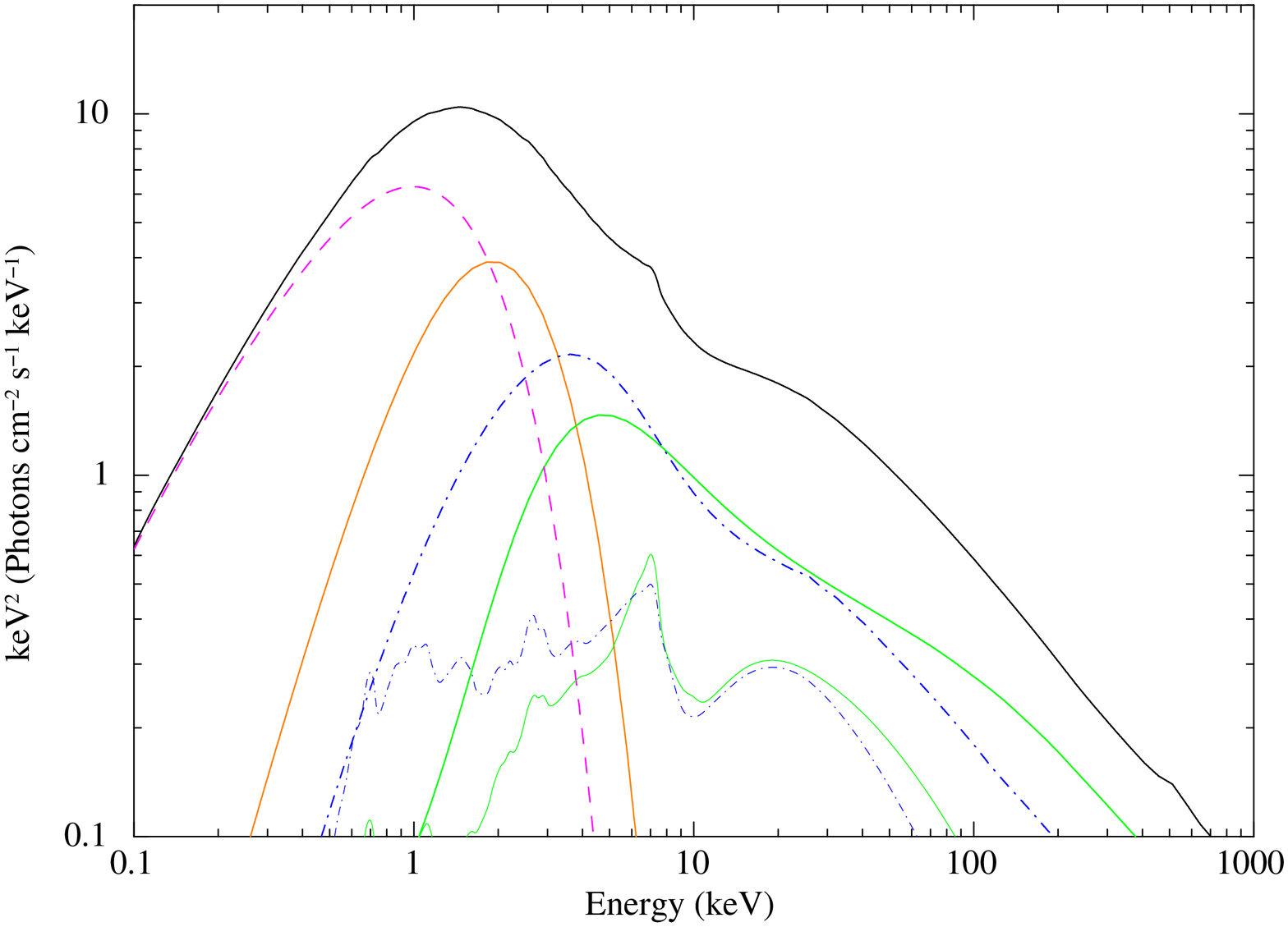}
 \caption{Unabsorbed spectral models for the Suzaku VHS spectrum based
 on best fit model of {\sc diskEQ} without inner flow (left), with
 inefficient untruncated inner flow (middle), and with stress-free
 truncated inner flow (right). Model components are the outer disk
 (dashed magenta), inner disk undernearth the coupled
 disc corona region (solid orange),
coupled hybrid corona and its reflection (solid green), 
inner flow and its reflection (dashed-dotted blue lines).}
\label{fig:model_dummy}
\end{figure*}

We again use the spin parameter to model the effect of a stress free
inner boundary condition at $R>R_{\rm ISCO}$. We again find a good
solution ($\chi^2/dof=183.6/199$) with $a^*=-1$ i.e. implying stress free
emissivity with truncation radius of $9R_g$ where the inner region is
now radiatively efficient (as there is much less energy), with
$f_{\rm rad}=1$.  The $\nu F_\nu$ spectrum deconvolved with this model is shown in 
Figure~\ref{fig:vhs-diskEQ} (panel a), while the residuals are 
shown in panel (d). The unabsorbed model components are shown in
the right panel of Figure~\ref{fig:model_dummy}, and 
best fit parameters are listed in column (4) in Table~\ref{tab:vhs}.
These show the 
coupled disc-corona region extends from 19--46~$R_{\rm g}$, while the
innermost flow extends from $\sim 9R_{\rm g}$ from $19R_{\rm g}$.

\subsection{The importance of reflection modeling}

The results shown in section 5.3 and 5.4 indicate that the efficiency
of the accretion disc decreases below the expected NT efficiency. One
of the scenarios to explain this inefficient inner flow is that a
fraction of the accretion power is used instead to power the jet.  The
energy loss inferred from the central flow is of order the radiated
luminosity. Thus we need to determine the coronal power to better than
a factor of 2 to constrain this energy loss. 
However, reflection can contribute to the hard X-rays
with luminosity of the same order as the accretion power, so treating
reflection correctly is important.

We can get an upper limit to the coronal power by removing reflection,
and just using an apparent {\sc smedge} and gaussian to get the
spectrum to fit.  We checked our results on the coupled disc-corona
model with inefficient/efficient hot inner flow (section 5.3, 5.4).
In the case of the inefficient untruncated inner flow (see section 5.3
and column (3) in Table~\ref{tab:vhs}), the radiative efficiency of
the inner flow is found to be larger, but still significantly below
unity, with $f_{\rm
  rad}=0.42^{+0.01}_{-0.06}$ with $\ell_{\rm h}/\ell_{\rm s}=0.71^{+0.10}_{-0.08}$
($g=0.42\pm0.03$) under the similar disc-corona geometry with $R_{\rm
  cor}=42\pm 4R_{\rm g}$ and $R_{\rm flow}=28^{+6}_{-5}R_{\rm g}$.
Similarly, in the case of the efficient truncated inner flow (section
5.4, column (4) in Table~\ref{tab:vhs}), the inner radius of the
efficient hot inner flow is estimated to be $R_{\rm
  in}=8.6^{+0.6}_{-0.5} R{\rm g}$ with
$\ell_{\rm h}/\ell_{\rm s}=0.69^{+0.12}_{-0.08}$ ($g=0.41^{+0.04}_{-0.03}$).
The larger radiated power required by removing reflection means
that  $R_{\rm in}$ becomes smaller, but it is still strongly required
to be truncated. 
Thus even allowing no contribution from reflection, the radiated power
inferred is still significantly less than that required by the NT
emissivity.

Conversely, the maximum reflection will give the minimum radiated
power, and hence the maximum decrease in efficiency from NT. We
neglected reflection from the overlap coupled disc-corona region which
is acceptable only if the reflector has such high ionisation parameter
that it is completely reflective (as assumed) or that the optical
depth of the corona is large. Given that the optical depth of our
corona is not large, at $\tau\sim0.46$ (see column (3) and (4) in
Table~\ref{tab:vhs}), we explore whether we can increase the
contribution of reflection by incorporating reflection from the
coupled disc-corona region. Hence
we refit the data based on the model described in section 5.3 and 5.4
by changing the inner radius of the smeared reflection from $R_{\rm
  flow}$ to $R_{\rm cor}$.  Slightly worse fits are obtained as
$\chi^2/dof=188.1/199$ and $189.2/199$ for the inefficient and
efficient inner flow, respectively.  However, the values of the
efficiency and the truncated radius are obtained as $f_{\rm
  rad}=0.27\pm 0.03$ and $R_{\rm in}=8.9^{+0.7}_{-0.2}R_{\rm g}$ ,
respectively, the values which are the same as previous fits within
90\% uncertainties (see column (3)(4) in Table~\ref{tab:vhs}).  

\section{Discussion}

\subsection{Summary of the results}

Fitting physical models for the complex spectra seen in the VHS offers
new insight into the properties of this complex state.  Our model
offers two main improvements on earlier attempts: firstly it models
the Comptonised emission with the {\sc eqpair} code, which self
consistently calculates the electron distribution from non-thermal
acceleration. This is not a power law, as assumed in {\sc simpl}
Comptonisation, but is hybrid as the low energy electrons thermalise
by Coulomb rather than Compton cooling. This has a profound impact on
the derived reflection parameters. Rather than requiring that the
source be reflection dominated, and that the reflection is strongly
smeared, the {\sc eqpair} models (which fit the data better than the
{\sc simpl} models) have $\Omega/2\pi=0.8$ and only moderate smearing
which does not constrain black hole spin. We caution that spin
determinations from reflection parameters are not always robust to
changes in the continuum model.

The second improvement in our code is that it specifically sets the
available energy in terms of the physical black hole parameters,
assuming a Novikov-Thorne emissivity. This is clearly appropriate for
the high/soft state taken a few days after our VHS observation.  A
comparison of these two spectra using the same physical black hole
parameters is especially compelling as they have similar
luminosities. The high/soft state is disc dominated, so the emissivity
is dissipated as (colour temperature corrected) blackbody emission at
each radius.  A blackbody is the most efficient radiator so a spectrum
with the same luminosity cannot have a mean photon energy which is
lower than this. Since the VHS spectrum is steep, this means that the
similar luminosity VHS spectrum cannot peak at a lower energy if the
dissipation as a function of radius remains unaltered. The observed
drop in energy at which the VHS spectrum peaks compared that of the
thermal high/soft state at a similar luminosity then clearly shows
that the energy dissipated in the accretion flow as a function of
radius has changed dramatically from that of Novikov-Thorne. Simply
growing a corona, even one whose energy and seed photons are coupled
to the underlying disc emission, will produce a spectrum which peaks
at higher energy than that of the multi-blackbody disc seen in the
high/soft state. Even making a transition from the coupled disc-corona
to a hot inner flow still cannot fit the data unless the dissipation
changes, fundamentally because a blackbody is the most efficient
radiator, so any other radiation mechanism will only increase the peak
spectral energy for the same luminosity.

There are three possible ways to match what is observed. Firstly the 
total dissipation could still be Novikov-Thorne if most of the power 
released in the innermost radii does not go to heating the accretion
flow but is instead 'dark', e.g. lost as kinetic energy in a
jet/wind. Secondly, the dissipation could be changed from
Novikov-Thorne in the inner regions e.g. if the flow is misaligned and
truncation at the bending wave radius sets an effective 
stress-free inner boundary to the flow at $R>R_{\rm ISCO}$. 
Both these require that the radiated efficiency has dropped compared
to Novikov-Thorne, so the same luminosity as the high/soft state can
only be produced with a higher mass accretion rate through the outer
(standard) disc. The third possibility is that the
flow is not in steady state, so that the mass accretion rate in
outer disc is larger than that in the inner disc. 
We discuss each of these in turn below.

\subsection{Low radiative efficiency: jet power?}

The most obvious solution to the observed discrepancy in luminosity
between the inner and outer flow is that the additional power is lost
up the jet.  The kinetic power of the jet is poorly constrained, but
may be of order the accretion power (Fender et al.~2004). This
would imply that the jet is accretion powered, rather than tapping the
spin energy of the black hole (Blandford-Znajek).

The source here is still showing type C QPOs, so has not made a HIMS/SIMS
transition which is (loosely) associated with the collapse of the radio jet, so 
on timing signatures alone the source could still be powering a steady, compact jet. 
However, there are simultaneous radio observations which show that the
jet is already strongly suppressed by this point (Corbel: private
communication). The 3--9~keV observed X-ray flux of $7.9\times
10^{-9}~{\rm ergs~s^{-1}cm^{-2}}$ predicts a 9~GHz radio flux of 15~mJy
from the fundamental plane \citep{corbel13_universal}. Removing the disc
component so as to include only the Comptonised power still gives a
3--9~keV flux of $7.4\times 10^{-9}~{\rm ergs~s^{-1}cm^{-2}}$, so this
does not make a significant difference to the predicted radio
flux. Yet the observed 9~GHz flux on 15th February (simultaneous with
the end of the Suzaku observation) is 1.4~mJy (Corbel, private
communication), so it appears that the radio is already strongly
suppressed below the fundamental plane at this point.  
Thus our missing luminosity is not
powering the steady, compact jet.
We note that a similar drop in radio without being accompanied by any change
in X-ray variability power and QPO type is also seen in MAXI~J$1659-152$\citep{horst13}.

Instead, the timing transition may be associated with the launching of the 
discrete ejection events \citep{fender09,miller-jones12}, which gives another
potential energy/mass loss process. 
However, our VHS observation
is still in the hard intermediate state, with strong X-ray variability
and type C QPO.  The transition to the SIMS happens  up to a day
after our VHS observation \citep{motta11}, and indeed, the radio
data just before and just after the high/soft state on the 19th
February show an optically thin radio flare (16 and 12~mJy at 9~GHz on
18 and 20th February).  Thus if the HIMS/SIMS transition marks the ejection, our
data are not losing energy via this route as the transition has not yet taken place. 

However, we note that radio behaviour around the transition is complex, and that there 
are time lags between various components of the jet which can make it difficult to 
establish causal connections (Corbel
et al 2013a; Miller-Jones et al 2012; Curran et al 2015;
Brocksopp et al 2013; Miller-Jones et al 2012; Hannikainen et al 2009).
Nonetheless,  it seems that the power is not being dissipated in a steady jet,
nor in a discrete ejection. Our Suzaku data are taken over a timescale
of 3 days without much change in spectral or timing properties (T12;
Motta et al 2011), so it seems unlikely that there was a missed
transition and then recovery back to the hard intermediate state.

There do not appear to be any other obvious energy loss channels.
The inferred mass accretion rate (modulo the uncertainties in system
parameters) does not seem to be at or above Eddington, so optically
thick advection (Abramowicz et al 1989), or mass loss through
radiatively driven winds (Shakura \& Sunyaev 1973) are not likely to
make much of a difference to the radiated power, though magnetic winds
from the inner disc are always a possibility. Conversely, the
system luminosity is high enough that substantial losses through
optically thin advection are not likely either (Narayan \& Yi 1996;
Yuan et al 2004). Thus a magnetically driven outflow/jet appears to be the
only candidate, and this needs to carry away around as much energy as
is radiated. 

\subsection{A stress-free inner boundary at $R>R_{\rm ISCO}$}

The Novikov-Thorne emissivity is specifically derived for a thin disc,
where the stresses which give rise to viscosity are in a plane. A
geometrically thick flow can have very different stresses e.g. 
a large scale height flow can have magnetic connections across the horizon, so
the stress-free inner boundary condition is no longer appropriate
\citep{krolik05,shafee08}.  However, this leads to
additional dissipation, rather than to less dissipation, as required
here.

Our preferred flow model is complex, with a thin disc, transitioning
to a coupled disc-corona region. The thin disc then disappears
completely in the inner region, leaving only corona plasma between
$R_{\rm flow}$ and $R>R_{\rm ISCO}$.  It is not at all clear
what happens to the dissipation at the edge of the thin disc $R_{\rm flow}$. The hotter plasma may support stresses like the
Novikov-Thorne stresses across the boundary between them, or it may be
rarified enough that a stress-free inner boundary condition at the
truncation radius is more appropriate. In the latter case, the
accretion power on the inner edge of the disc-corona and the outer part of the hot flow
is lower than expected
from Novikov-Thorne, though this might not be a large enough effect to 
produce the required marked deficit in power.

Instead, the mismatch between mass accretion rate and luminosity in
the inner flow could be produced in the three zone models as an
additional signature of a Lense-Thirring origin of the QPO. A flow
which is misaligned to the black hole spin experiences a torque, which
propagates at the local sound speed.  A thin disc is cool, so the
bending waves cannot propagate fast enough to make the disc precess,
so it forms a stable warp (Bardeen-Petterson).  Conversely, for a
thick disc, the hotter material means a faster sound speed, so the
torque can communicate across the flow on timescales fast enough for
it to precess as a solid body (Fragile et al 2007; Ingram et al 2009).
The misalignment torques are so strong at small radii that they
effectively truncate the flow at a bending wave radius, of order
8--12$R_g$ (Fragile et al 2007; 2009; Ingram et al 2009; Dexter \& Fragile 2011; Ingram \&
Done 2012). The accretion flow is laminar rather than turbulent after
this point, so it may lead to an effective stress--free inner boundary
condition at a radius larger than $R_{\rm ISCO}$, giving less accretion
power. However, the torques could also simply transfer the accretion
power from below to above the bending wave radius, so that the total
dissipation in the flow remained similar to that expected from
Novikov-Thorne (Fragile,
private communication).

\subsection{Applicability of Novikov-Thorne emissivity}

The two possibilities discussed above both require that the overall
efficiency is lower than Novikov-Thorne. Fundamentally this conclusion is
driven by the observation that the VHS 
spectrum peaks at lower energy than that seen in the similar luminosity
high/soft state. The observed continuum below 1~keV is sensitive to
the mass accretion rate through the outer (standard disc).  The
Novikov-Thorne emissivity should be appropriate here, yet fits
constrain the mass accretion rate through the outer disc to be $\sim
1.5\times$ higher than that required to power the high/soft state (see
Tables~\ref{tab:hs} and \ref{tab:vhs}). Since the observed luminosity
of the VHS is only $\sim 1.1\times $ that of the  high/soft state, then this requires that the
radiative efficiency is lower in the inner disc. 

However, a decreasing mass accretion rate from the VHS to the
high/soft state while the bolometric luminosity stays almost the same
appears somewhat fine-tuned. 
Instead, since our data are taken on the fast rise to outburst, the
flow could simply be out of steady state, while Novikov-Thorne
explicitly assumes constant $\dot{M}$ at all radii.  Simulations of
the viscous evolution of the disc during an outburst show that the
surface density takes days to settle into this configuration after the
the Hydrogen ionisation instability is triggered at large radii \citep{dubus01}. 
 Nonetheless, our VHS data are taken over a 3 day time
period with little evolution in spectra or timing properties, so it
seems more likely that the flow is in some sort of steady state.

\subsection{Distinguishing between these possibilities}

The VHS is a rather rare spectral state. More observations would
clearly show whether the difference in emissivity seen here is only
associated with fast changes in mass accretion rate, as is required in section 6.4. 
Similarly, more simultaneous radio observations would tie down
the associated jet and/or plasma ejections required in section 6.2, though the
magnetic wind is much harder to constrain.  A stress-free inner
boundary due to bending wave radius from a misaligned precessing hot
flow (section 6.3) should also be present in any QPO observation of the low/hard
state.

\section{Conclusions}

Comptonisation in the VHS is complex, with a high energy tail pointing
to the importance of non-thermal electrons. However, purely
non-thermal acceleration does not result in a purely power law
electron distribution when both Compton and Coulomb collisions are
included. The self-consistent (pure power law acceleration balancing
cooling) steady state electron distribution is thermal at low
energies, resulting in a Compton continuum with complex curvature, as
calculated in the {\sc eqpair} code by Coppi. This
completely removes the requirement for high black hole spin and high
reflected fraction in the Suzaku VHS data.

We develop a new model for the accretion flow, {\sc diskEQ} which
tracks energy across the flow but allows this to be dissipated either
in a standard disc, or in a coupled disc-corona geometry where the
Comptonisation is based on the {\sc eqpair} code or in a hot inner
flow, where again the Comptonisation is described by {\sc eqpair}.
Using this, we demonstrate for the first time that the VHS is not well
described by Novikov-Thorne emissivity. The spectrum below 1~keV
constrains the outer (standard) disc emission, requiring a mass
accretion rate which is $1.5\times $ higher than that seen in a disc
dominated high/soft state, but the total VHS spectrum is only a factor
$1.1\times $ brighter. The most obvious solution is that the excess
Novikov-Thorne emissivity in the inner disc is not translated into
radiation, but is instead used to power of the jet. However,
simultaneous observations show that the radio is strongly suppressed.
Instead, it could show that the Novikov-Thorne prescription is not
appropriate, perhaps due to the flow being misaligned as required to 
produce the QPO from Lense-Thirring precession. 

We urge the use of more physical models to fit the complex
intermediate states in order to distinguish between these
possibilities, and are releasing the code within the {\sc xspec}
spectral fitting package to enable this.

\appendix
\section{Approximation of the truncated radius with friction free inner boundary condition}
We allow the distance $d$ to be
an additional free parameter in order for it to act as a proxy
for the truncation radius of a flow/disc with stress free inner boundary at $R>R_{ISCO}$, 
for the fixed spin parameter $a^\ast=-1$.
The scaling factor in Table~\ref{tab:vhs} is defined as $(d/8~{\rm kpc})^{-1}$, 
and the proper radii are approximately estimated by multiplying the scaling factor to the derived radii  with {\sc diskEQ}.   

\section{The impact of reprocessing/reflection on $\ell_{h}/\ell_{s}$ in the coupled disc-corona}

The ratio $\ell_{\rm s}/\ell_{\rm h}$ is the major parameter determines the
spectral shape from Comptonisation. In the overlap region, reflection
and reprocessing mean that this is not simply related to the fraction
$f$ of the intrinsic power which is dissipated in the corona. At the
disc surface, the outgoing flux $\ell_{\rm disk}^{\rm  out}$ is the sum of the intrinsic disc
dissipation, $\ell_{\rm si}$, and the reflected/reprocessed response of
the disc to the irradiating flux from the corona, $\ell_{\rm rep}$, and reflected component, $\ell_{\rm R}$.
Hence $\ell_{\rm disk}^{\rm  out}=\ell_{\rm si}+\ell_{\rm R}+\ell_{\rm rep}=\ell_{\rm s}+\ell_{\rm R}$ as the
total soft flux is the sum of the seed and reprocessed emission. 
By using disc irradiation from the corona, $\ell_{\rm ir}$,
$\ell_{\rm R}$ can be written as $\ell_{\rm R}=a\ell_{\rm ir}$ for an albedo $a$,
 and $\ell_{\rm ir}$ is the 
sum of downwards flux from intrinsic dissipation, plus the energy of
the seed photons backscattered from the disc. 
Hence,
considering the fraction of disc and reflection luminosity scattered in the corona, $\ell_{\rm scat}=(\ell_{\rm s}+\ell_{\rm R})(1-e^{-\tau})$, 
the irradiation from the corona is written as following.
\[
 \ell_{\rm ir}=\ell_{\rm h}/2 +
(1-e^{-\tau})\ell_{\rm disk}^{\rm out}/2 = \ell_{\rm h}/2 +
(1-e^{-\tau})(\ell_{\rm s}+a\ell_{\rm ir})/2
\]
Recasting this as an explicit equation gives 
\[
\frac{\ell_{\rm ir}}{\ell_{\rm h}}=\frac{1+(\ell_{\rm s}/\ell_{\rm h})(1-e^{-\tau})}{2-a+ae^{-\tau}}=\frac{e^{-\tau}  + (1-e^{-\tau})/g}{2-a+ae^{-\tau}} 
\]
defining $g=\ell_{\rm h}/(\ell_{\rm h}+\ell_{\rm s})$, similar to the equation
$f=\ell_{\rm h}/(\ell_{\rm h}+\ell_{\rm si})$ for the intrinsic dissipation.
Since $\ell_{\rm si}=\ell_{\rm s}-\ell_{\rm rep}$ then 
$f=g(1-g (1-a)\ell_{\rm ir}/\ell_{\rm h})^{-1}$. Substituting from above for $\ell_{\rm ir}/\ell_{\rm h}$ gives
\[
f=g\frac{2-a+ae^{-\tau}}{1+e^{-\tau} - g(1-a) e^{-\tau}}
\]
(J. Malzac, private communication).  

The power emitted from the top of the corona, $\ell_{\rm out}$,
is now the sum of the coronal power plus the seed photons travelling
upwards, $(\ell_{\rm h}+\ell_{\rm scatt})/2$, together with the disc
emission crossing the corona without any interaction, $(\ell_{\rm
  s}+\ell_{\rm R})e^{-\tau}$. The total disc emission is caused by
internal dissipation and irradiation so $\ell_{\rm s}+\ell_{\rm
  R}=\ell_{\rm si}+\ell_{\rm ir}$. Substituting for $\ell_{\rm ir}$
from above gives $\ell_{\rm s}+\ell_{\rm R}=2(\ell_{\rm si}+\ell_{\rm
  h}/2)/(1+e^{-\tau})$ so that $\ell_{\rm out}=\ell_{\rm si}+\ell_{\rm
  h}$. Hence energy is still conserved in this region, as assumed in
the code (J. Malzac, private communication). However, our code does
not explicitly distinguish the internally reflected emission, $\ell_R$,
from $\ell_h$, and we caution that full radiative transfer is probably
required in this region. 

\subsection*{Acknowledgements}

We are grateful to Julien Malzac as a referee for his valuable
  and helpful comments and suggestions, especially for the derivation
  in Appendix B.
We thank Mari Kolehmainen for giving us  the high/soft state spectrum of GX~$339-4$, and Stephane Corbel for giving us a radio data.
We also thank Shin'ya Yamada for his helpful advice on GSO analyses, and Chris Fragile for
insight into the simulations. 
AK and CD acknowledge support from Grant-in-Aid No.24540237 from Ministry of Education, Culture, Sports, Science
and Technology of Japan. 
CD also acknowledges support from the STFC consolidated grant ST/L00075X/1.

\end{document}